\documentclass{emulateapj}

\shorttitle{Evolution of radio selected AGN}
\shortauthors{V. Smol\v{c}i\'{c} et al.}

\def\f#1   {Fig.~\ref{#1}}
\def\s#1   {Sec.~\ref{#1}}
\def\tab#1   {Tab.~\ref{#1}}
\def\t#1   {Tab.~\ref{#1}}
\def\lum   {$\mathrm{L}_\mathrm{1.4GHz}$}
\def\comm#1   {{\tt (COMMENT: #1) }}

\def\sqdegs           {$\Box^{\circ}$}

\def\msol              {$\mathrm{M}_{\odot}$}

\def\wh                {W~Hz$^{-1}$}
\def\whs                {W~Hz$^{-1}$~sr$^{-1}$}

\def\Vm                {$V_\mathrm{max}$}

\slugcomment{  }

\begin{document}

\title{ Cosmic evolution of radio selected active galactic nuclei in
the COSMOS field\altaffilmark{0}}

\author{V.~Smol\v{c}i\'{c}\altaffilmark{1},
        G.~Zamorani\altaffilmark{2},
        E.~Schinnerer\altaffilmark{3},
        S.~Bardelli\altaffilmark{2},
        M.~Bondi\altaffilmark{4},
        L.~B$\hat{i}$rzan\altaffilmark{5,6}, 
        C.~L.~Carilli\altaffilmark{7},
        P.~Ciliegi\altaffilmark{2},
        O.~Ilbert\altaffilmark{8},
	A.~M.~Koekemoer\altaffilmark{9},
        A.~Merloni\altaffilmark{10,11}
        T.~Paglione\altaffilmark{12,13}, 
        M.~Salvato\altaffilmark{1},
        M.~Scodeggio\altaffilmark{14},
        N.~Scoville\altaffilmark{1}
        }
\altaffiltext{0}{Based on observations with the National
Radio Astronomy Observatory which is a facility of the National Science
Foundation operated under cooperative agreement by Associated Universities,
Inc. }
\altaffiltext{1}{ California Institute of Technology, MC 105-24, 1200 East
California Boulevard, Pasadena, CA 91125 }
\altaffiltext{2}{INAF - Osservatorio Astronomico di Bologna, via Ranzani 1,
  40127, Bologna, Italy} 
\altaffiltext{3}{ Max Planck Institut f\"ur Astronomie, K\"onigstuhl 17,
  Heidelberg, D-69117, Germany } 
\altaffiltext{4}{INAF - Istituto di Radioastronomia, via Gobetti 101, 40129
  Bologna, Italy }
\altaffiltext{5}{Department of Astronomy and Astrophysics, Pennsylvania State University, University Park, PA 16802}
\altaffiltext{6}{Department of Physics and Astronomy, Ohio University, Clippinger Laboratories, Athens, OH 45701}
\altaffiltext{7}{National Radio Astronomy Observatory, P.O. Box 0, Socorro,
  NM 87801-0387 } 
\altaffiltext{8}{ Institute for Astronomy, 2680 Woodlawn Dr., University of
  Hawaii, Honolulu, Hawaii, 96822 }
\altaffiltext{9}{Space Telescope Science Institute, 3700 San Martin Drive, Baltimore, MD 21218}
\altaffiltext{10}{Excellence Cluster Universe, Technische Universität München, 
Boltzmannstr. 2, D-85748, Garching, Germany}
\altaffiltext{11}{Max-Planck-Institut f\"{u}r Extraterrestrische Physik, Giessenbachstr., D-85741 Garching, Germany}
\altaffiltext{12}{York College, City University of New York, 94-20 Guy R. Brewer
  Boulevard, Jamaica, NY 11451} 
\altaffiltext{13}{American Museum of Natural History,
Central Park West at 79th Street, New York, NY 10024} 
\altaffiltext{14}{IASF Milano-INAF, Via Bassini 15, I-20133, Milan, Italy }

\begin{abstract}
  We explore the cosmic evolution of radio AGN with low radio powers
  (\lum~$\lesssim5\times10^{25}$~\wh ) out to $z=1.3$ using to-date
  the largest sample of $\sim600$ low luminosity radio AGN at
  intermediate redshift drawn from the VLA-COSMOS survey.  We derive
  the radio luminosity function for these AGN, and its evolution with
  cosmic time assuming two extreme cases: i) pure luminosity and ii)
  pure density evolution. The former and latter yield $L_*\propto
  (1+z)^{0.8\pm0.1}$, and $\Phi_*\propto (1+z)^{1.1\pm0.1}$,
  respectively, both implying a fairly modest change in properties of
  low radio-power AGN since $z=1.3$.  We show that this is in stark
  contrast with the evolution of powerful
  (\lum~$>5\times10^{25}$~\wh ) radio AGN over the same cosmic
  time interval, constrained using the 3CRR, 6CE, and 7CRS radio
  surveys by \citet{willott01}.  We demonstrate that this can be
  explained through differences in black hole fueling and triggering
  mechanisms, and a dichotomy in host galaxy properties of weak and
  powerful AGN. Our findings suggest that high and low radio-power AGN
  activity is triggered in different stages during the formation of
  massive red galaxies. We show that weak radio AGN occur in the most
  massive galaxies already at $z\sim1$, and they may significantly
  contribute to the heating of their surrounding medium and thus
  inhibit gas accretion onto their host galaxies, as recently
  suggested for the `radio mode' in cosmological models.
\end{abstract}


\section{Introduction}
\label{sec:introduction}

\subsection{ AGN feedback: The impact of radio sources on galaxy formation  }

Radio activity from active galactic nuclei (AGN) has recently been
invoked in cosmological models as a significant ingredient in the
process of galaxy formation (`AGN feedback'). Given that, in the past,
cosmological models led to a systematic over-prediction of the
high-mass end of the galaxy stellar mass function
\citep[e.g.][]{white78,white91}, the implementation of gas heating
through energetic radio outflows -- the `radio' mode\footnote{The
  `radio mode' is defined here as given in \citet{croton06}.} --
managed to reproduce well many observed galaxy properties (e.g.\ the
galaxies' stellar mass function, color bimodality; Croton et al. 2006,
Bower et al.  2006).  The particular choice of a `radio mode' as the
relevant heating mechanism has been motivated by many observational
results verifying the interplay between the emission of radio galaxies
in galaxy clusters, and the cluster's hot X-ray emitting gas on
large-scales (e.g.\ Fabian et al.\ 2003, Forman et al.\ 2005).

In the centers of galaxy clusters the radiative cooling time of the
intra-cluster medium (ICM) is shorter than the age of the
cluster. Thus, the central ICM is expected to be significantly colder
compared to outskirt-regions (such clusters are referred to as
`cooling flow clusters'). However, generally the expected cool X-ray
phases are not observed, and this is known as the `cooling flow
problem'.  The most likely solution to this problem is thought to be
radio galaxies as it is usually observed that the synchrotron plasma
ejected by their radio jets inflates bubbles in the hot X-ray emitting
cluster gas and thereby heats it (see \citealt{fabian94} for a
review). In theoretical models, a similar interplay on smaller scales
has been assumed to be at work between the radio outflows of a galaxy
and its surrounding hot gas halo.

The first observational evidence for the `radio mode' feedback in the
context of galaxy formation has been found by \citet{best06}.  Based
on X-ray and radio observations of a local sample of massive
elliptical galaxies \citep{best05} they have shown that the radio
source heating may indeed balance the radiative energy losses from the
hot gas surrounding the galaxy, as postulated in the models.  Based on
an independent sample of radio galaxies in galaxy clusters (studied by
\citealt{birzan04}) \citet{best06} have correlated the mechanical
heating provided by the radio sources on their surrounding medium with
monochromatic 20~cm radio power. Combining this with the local radio
AGN luminosity function \citep{best05}, they have found that the
volume-averaged mechanical energy heating rate of local radio luminous
AGN is about a factor of 10-20 lower than predicted by the Croton et
al.\ (2006) model.  They assigned this difference to a systematic
over-prediction of the heating rate in the model (by a factor of 2-3)
and the intention of the model to jointly describe both the large- and
small- scale heating from radio sources (i.e.\ on both cluster and
galaxy scales). More recent calculations of the volume averaged
mechanical energy heating rate due to AGN \citep{koerding08,
merloni08} yield higher values at $z=0$ compared to the results of
\citet{best06}, and thus closer to the \citet{croton06} prediction. 

To date no clear picture exists on how the AGN radio mode feedback
works.  More observations are needed to provide better constraints on
the physics of this process, as well as its evolution with cosmic
time. Here we attempt to shed light on the latter utilizing a large
statistical sample of radio AGN, drawn from the VLA-COSMOS survey
\citep{schinnerer07}.

\subsection{ Radio luminous AGN and their evolution }

\citet[FR hereafter]{fr74} were the first to note that the radio
luminous AGN population consists of two apparently distinct types of
sources, referred to as FR type I and II, with prominent differences
in both radio morphology and luminosity. The radio emission from FR~I
radio galaxies is core dominated, while FR~IIs are edge-brightened
with highly collimated large-scale jets.  FR~IIs are typically more
powerful in the radio than FR~Is, and a dividing luminosity of
$L_\mathrm{178MHz} \sim 2.5\times10^{26}$~\wh\ (corresponding to
$L_\mathrm{1.4GHz}\sim6\times10^{25}$~\wh )\footnote{A spectral index
  of $\alpha=0.7$ ($F_\nu\propto\nu^{-\alpha}$; where $F_\nu$ is the
  radio flux density) is assumed throughout the paper.} has been
suggested \citep{fr74}. It has been later shown \citep{ledlow96} that
the FR~I/FR~II division is also a function of the host galaxy optical
luminosity (with a higher dividing radio luminosity for higher optical
host luminosity).

An alternative way of classifying radio AGN is based on the existence
of high excitation (HE) emission lines in the optical spectra of their
host galaxies \citep{hine79, laing94}. In this scheme, objects without
high-excitation emission lines are referred to as low-excitation (LE)
radio galaxies, and they are most common at low radio
luminosities. Almost all FR~I radio galaxies are LE sources, while
optical hosts of the most powerful radio sources, i.e.\ FR~IIs,
usually have strong emission lines. It is noteworthy, however, that
the correspondence between the FR class and the presence of emission
lines is not one-to-one.  Many FR~II galaxies have been found to be
low-excitation radio galaxies \citep[e.g.\ ][]{evans06}.

Recently, strong evidence \citep{evans06,hardcastle06,hardcastle07}
has emerged supporting the idea that low- and high- excitation radio
AGN (hereafter LERAGN and HERAGN, respectively) reflect different
modes of black hole (BH) accretion. Independent studies have shown
that i) LERAGN are a class of radio luminous AGN that accrete
radiatively inefficiently \citep{evans06, hardcastle06}, and ii) Bondi
accretion of the hot X-ray emitting inter-galactic medium (IGM) is
sufficient to power the jets of low-power radio galaxies in the
centers of galaxy clusters (based on a sample of nearby galaxies;
Allen et al.\ 2006). Based on these findings Hardcastle et al.\ (2007)
have developed a model in which all low-excitation radio sources are
powered by radiatively inefficient accretion (i.e.\ at sub-Eddington
accretion rates) of the hot phase of the IGM, while high-excitation
sources are powered by radiatively efficient accretion (at
$\sim$~Eddington rates) of cold gas (that is in general unrelated to
the hot IGM; see also \citealt{merloni08}). This model successfully
explains a variety of properties of radio luminous AGN, such as their
environments, signs of galaxy mergers in the hosts of powerful
(high-excitation) radio sources, and the break of the radio AGN
luminosity function (for details see \citealt{hardcastle07} and
references therein).

In the past two decades it has become clear that radio luminous AGN
evolve differentially: Low-power sources evolve less strongly than
high-power sources. Numerous studies of high luminosity radio AGN
\citep[$L_\mathrm{2.7GHz}\gtrsim10^{25}$~\whs\ corresponding to
  \lum~$\sim2\times10^{26}$~\wh ;][]{dp90, willott01} have found a
strong positive density evolution with redshift of these sources out
to a redshift peak of $z\sim2$. Beyond this redshift their comoving
volume density starts declining (a possible sharp density cut-off has
been suggested by \citealt{dp90}, but has not been confirmed by
\citealt{willott01}). Such a decline would be consistent with the
results obtained via optical surveys \citep{schmidt95} and, recently,
X-ray surveys \citep{silverman08, brusa08}.

Analyzing the evolution of lower-power radio AGN
(\lum~$>2\times10^{25}$~\wh ) \citet{waddington01} have found a
significantly slower evolution of this population, with the comoving
volume density turn-over occurring at a lower redshift ($z\sim1-1.5$),
compared to the high-power radio population.  %In general, in the low
%luminosity regime( \lum~$\sim10^{24-27}$~\wh ) a principally weaker
%evolution has been found compared to the high radio-power
%AGN.  However, the results for lower-power radio AGN, based on
Different radio-optical surveys are still somewhat
controversial. While \citet{clewley04} and \citet{sadler07} find no
evidence for particularly {\em strong} evolution out to $z\sim0.7$,
\citet{cowie04} have found a strong density evolution out to
$z\sim1.5$.  Thus, although on average a weaker evolution of low-power
(compared to high-power) radio AGN is implied, it is still not very
well understood how the low-luminosity radio AGN evolve with redshift.
In this work we use the VLA-COSMOS AGN sample of low luminosity
($96\%$ have \lum~$<10^{25}$~\wh ) radio sources in order to
comprehensively constrain the evolution of the low-power radio AGN out
to $z=1.3$. We combine our results with the new findings and ideas on
the cosmological relevance of radio AGN in order to study the impact
of radio luminous AGN on galaxy formation.

The paper is outlined as follows. In \s{sec:sample} \ we define the
VLA-COSMOS AGN sample. In \s{sec:lf} \ we derive the radio luminosity
function for the VLA-COSMOS AGN, and extend it to high radio powers
using the results from \citet{willott01}. In \s{sec:lfevolv} \ we
constrain the evolution of radio AGN out to $z=1.3$. In \s{sec:props}
\ we analyze the properties of local and intermediate redshift weak
and powerful radio AGN; in \s{sec:sfagntrigrates} \ we compute the
radio AGN mass function, and derive and compare the star formation
quenching and radio-AGN triggering rates. We discuss our results in
the context of galaxy formation and evolution in \s{sec:discussion} ,
and summarize them in \s{sec:summary} .

We report magnitudes in the AB system, adopt $H_0=70,\, \Omega_M=0.3,
\Omega_\Lambda = 0.7$, and define the radio synchrotron spectrum as
$F_{\nu} \varpropto \nu^{-\alpha}$, assuming $\alpha = 0.7$.

\section{ The VLA-COSMOS radio AGN sample } 
\label{sec:sample}

The sample of AGN galaxies used here is presented in
\citet[S08a hereafter]{smo08a}, and we briefly summarize it below.

Using radio and optical data for the COSMOS field, S08a have
constructed a sample of 601 AGN galaxies with $z\leq1.3$ from the
entire VLA-COSMOS Large Project catalog \citep{schinnerer07}. The
selection required optical counterparts with $i_\mathrm{AB}\leq26$,
accurate photometry, and a $\mathrm{S/N}\geq5$
(i.e.\ $\gtrsim50~\mathrm{\mu}$Jy) detection at 20~cm, and is based on
a rest-frame optical color classification (see also \citealt{smo06}).
The classification method was well-calibrated using a large local
sample of galaxies ($\sim7,000$ SDSS ``main'' spectroscopic galaxy
sample, NVSS%\footnote{National Radio Astronomy Observatory VLA Sky
%  Survey (NVSS; \citealt{condon98})} 
and IRAS
%\footnote{Infrared   Astronomical Satellite \citep{neugebauer84}} 
surveys) representative
of the VLA-COSMOS population. It was shown that the method agrees well
with other independent classification schemes based on mid-infrared
colors \citep{lacy04,stern05} and optical spectroscopy
(\citealt{bpt81,best05}). The selected sample of AGN is estimated to
be $\sim 90\%$ complete.

The rest-frame color classification procedure efficiently selects
mostly type 2 AGN (such as LINER, Seyferts), and absorption-line AGN
(with no emission lines in the optical spectrum), while type 1 AGN
(i.e.\ quasars, $\lesssim20\%$ of the total AGN sample) are not
included in the current sample (see S08a for detailed definitions of
the samples).

Although based on a color identification (as opposed to an optical
spectroscopic classification), the output of the selection of our
intermediate redshift ($z\leq1.3$) AGN sample is comparable to those
of numerous local ($z\leq0.3$) radio AGN samples extensively studied
in the literature \citep[e.g.][drawn from the SDSS,
%\footnote{ Sloan     Digital Sky Survey (SDSS;
%    \citealt{york00,stoughton02,abaz03,abaz04,abaz05,adelmc06,adelmc07})},
  2dF, and 6dF
%\footnote{2dF: Two Degree Field; 6dF: Six Degree Field; {\em magnum.anu.edu.au/~TDFgg}} 
optical surveys combined with the
  NVSS and FIRST
%\footnote{Faint Images of the Radio Sky at
%    Twenty-centimetres (FIRST; \citealt{bec95})} 
20~cm radio
  surveys]{sadler02,best05,mauch07}. This is an important feature as
it enables a straight-forward and fair comparison of our results with
those based on these local studies (e.g.\ the local AGN radio
luminosity function).

Out of the 601 selected AGN galaxies 262 have spectroscopic redshifts,
while the remaining sources have very reliable photometric redshifts
available ($\sigma\left( \frac{\Delta z}{1+z} \right )=0.027$; see
S08a and references therein).  Based on Monte Carlo simulations, S08a
have shown that the photometric errors in the rest-frame color
introduce a small, $\sim5\%$, uncertainty in the number of the
selected galaxies.  Here we use the S08a sample of AGN galaxies,
statistically corrected for this effect.

\section{ The 1.4~GHz luminosity function for radio AGN  in
  VLA-COSMOS}
\label{sec:lf}

\subsection{Derivation of the luminosity function (LF)}

We derive the radio LF ($\Phi$) for our AGN galaxies in four redshift
bins using the standard $1/$\Vm\ method \citep{schmidt68}.  We limit
the accessible volumes a) on the bright end by the minimum
  redshift cut-off of the redshift range in consideration or the
minimum redshift out to which an object could be observed due to the
optical saturation limit of $i^*=16$ (AB mag; see also
\citealt{capak07}), and b) on the faint end by the maximum redshift
out to which a galaxy could be observed given the flux limits on both
the radio and optical data. In practice, the latter is dominated by
the radio detection limits, rather than the optical, as the major
fraction of the sources used here has $i_\mathrm{AB}$ band magnitudes
brighter than $\sim24$ (see Fig.~21 in S08a).

We further take into account the non-uniform rms noise level in the
VLA-COSMOS mosaic via the differential visibility area (i.e.\ areal
coverage, $A_k$, vs. rms; see Fig.~13 in
\citealt{schinnerer07}). Hence, for a source with a 1.4~GHz luminosity
$L_j$ its maximum volume is $V_\mathrm{max}(L_j) = \sum_{k=1}^{n} A_k
\cdot V_\mathrm{max}(z_\mathrm{max}^{A_k},L_j)$.

In order to robustly derive the LF we take several additional
corrections into account: a) the VLA-COSMOS detection completeness
\citep{bondi08}, and b) the AGN galaxy selection bias due to the
rest-frame color uncertainties. We correct for these in the same way
as described in \citet[S08b hereafter]{smo08b}. The median of the
first correction (as a function of flux density) is $\sim10\%$
(reaching a maximum value of 60\% in only one of the lowest flux
density bins; see Tab.~1 in \citealt{bondi08}). The second correction,
based on Monte Carlo simulations, yields a reduction of $\sim5\%$ of
the average number of radio AGN (see below; see also S08b for a more
detailed description).

Hence, in the i$^\mathrm{th}$ luminosity bin the
comoving space density ($\Phi_i$), and its corresponding error
($\sigma_i$), are computed by weighting the contribution of each
(j$^\mathrm{th}$) galaxy by the completeness correction factor,
$\mathrm{f_{det}}$ (see \citealt{bondi08}):
\begin{eqnarray} 
\Phi_i = \sum_{j=1}^{N} 
              \frac{\mathrm{f_{det}^j} }
                   {\mathrm{V_{max}^j}} \,\, ; \quad
\sigma_i = \sqrt{ \sum_{j=1}^{N} 
              \left( 
               \frac{\mathrm{f_{det}^j}  }
                   {\mathrm{V_{max}^j}} 
               \right)^2
              }
\end{eqnarray}

The selection bias due to the rest-frame color uncertainties is
accounted for via Monte Carlo simulations.  As described in S08a and
S08b, in each iteration the rest-frame color error distribution is
generated (see Fig.~5 in S08a) and added to the rest-frame color
derived by SED fitting. AGN galaxies are then re-selected, and the LF
is derived as described above.  In this way we obtain 100 realizations
of ($\Phi_i$,$\sigma_i$) for each luminosity bin, and we take the
median values as representative.

\subsection{The radio AGN luminosity function }

\begin{deluxetable}{c|c|c}
\tablecaption{ Luminosity functions for VLA-COSMOS AGN
\label{tab:LFs}
  }
\tablehead{
\colhead{ redshift} &
\colhead{$\mathrm{L_{1.4GHz}}$} &
\colhead{$\Phi $}\\
\colhead{range} &
\colhead{$[\mathrm{ W Hz^{-1}}]$} &
\colhead{$[\mathrm{Mpc^{-3} dex^{-1}]}$}
}
\startdata
                & $ 4.47\cdot10^{21}$ & $5.08 \pm 2.25\cdot10^{-4}$ \\
                & $ 1.41\cdot10^{22}$ & $2.54 \pm 0.66\cdot10^{-4}$ \\
                & $ 4.47\cdot10^{22}$ & $2.22 \pm 0.40\cdot10^{-4}$ \\
                & $ 1.58\cdot10^{23}$ & $4.54 \pm 1.57\cdot10^{-5}$ \\
$0.1<z\leq0.35$ & $ 5.62\cdot10^{23}$ & $1.21 \pm 0.70\cdot10^{-5}$ \\
                & $ 1.78\cdot10^{24}$ & $1.66 \pm 0.88\cdot10^{-5}$ \\
                & $ 5.62\cdot10^{24}$ & $7.90 \pm 5.59\cdot10^{-6}$ \\
                & $ 1.78\cdot10^{25}$ & $1.46 \pm 0.84\cdot10^{-5}$ \\
\hline
                & $ 3.16\cdot10^{22}$ & $1.90 \pm 0.55\cdot10^{-4}$ \\
                & $ 7.94\cdot10^{22}$ & $1.70 \pm 0.29\cdot10^{-4}$ \\
                & $ 2.00\cdot10^{23}$ & $4.48 \pm 1.18\cdot10^{-5}$ \\
$0.35<z\leq0.6$ & $ 5.62\cdot10^{23}$ & $1.40 \pm 0.57\cdot10^{-5}$ \\
                & $ 1.41\cdot10^{24}$ & $8.96 \pm 3.96\cdot10^{-6}$ \\
                & $ 3.16\cdot10^{24}$ & $4.49 \pm 2.81\cdot10^{-6}$ \\
                & $ 7.94\cdot10^{24}$ & $4.53 \pm 2.84\cdot10^{-6}$ \\
\hline
                & $ 1.00\cdot10^{23}$ & $8.76 \pm 2.32\cdot10^{-5}$ \\
                & $ 2.51\cdot10^{23}$ & $8.75 \pm 1.36\cdot10^{-5}$ \\
                & $ 6.31\cdot10^{23}$ & $2.86 \pm 0.61\cdot10^{-5}$ \\
$0.6<z\leq0.9$  & $ 1.78\cdot10^{24}$ & $2.45 \pm 0.50\cdot10^{-5}$ \\
                & $ 4.47\cdot10^{24}$ & $4.11 \pm 1.82\cdot10^{-6}$ \\
                & $ 1.00\cdot10^{25}$ & $3.43 \pm 1.66\cdot10^{-6}$ \\
                & $ 2.24\cdot10^{25}$ & $6.88 \pm 7.45\cdot10^{-7}$ \\
\hline
                & $ 3.16\cdot10^{23}$ & $3.72 \pm 0.93\cdot10^{-5}$ \\
                & $ 7.08\cdot10^{23}$ & $2.28 \pm 0.47\cdot10^{-5}$ \\
                & $ 1.78\cdot10^{24}$ & $1.82 \pm 0.34\cdot10^{-5}$ \\
$0.9<z\leq1.3$  & $ 5.01\cdot10^{24}$ & $3.97 \pm 1.38\cdot10^{-6}$ \\
                & $ 1.26\cdot10^{25}$ & $4.25 \pm 1.33\cdot10^{-6}$ \\
                & $ 3.16\cdot10^{25}$ & $1.42 \pm 0.77\cdot10^{-6}$ \\
                & $ 7.08\cdot10^{25}$ & $7.16 \pm 5.48\cdot10^{-7}$ \\
                & $ 1.78\cdot10^{26}$ & $1.46 \pm 0.79\cdot10^{-6}$ 
\enddata
\tablenotetext{.}{$\mathrm{H_0}=70,\, \Omega_M=0.3, \Omega_\Lambda = 0.7$}
\end{deluxetable}

The 1.4~GHz radio LFs for our AGN galaxies for the $4$ chosen redshift
bins are shown in \f{fig:lf} , and tabulated in \t{tab:LFs} .  In each
panel in \f{fig:lf} \ we also show the local 20~cm LFs for AGN derived
by \citet{condon02}, \citet{sadler02}, \citet{best05}, and
\citet{mauch07}.  Our derived LF in the local redshift bin (top left
panel) agrees well with the local LFs. This is quite remarkable as a)
our identification method is based only on photometry contrary to the
spectroscopic selection performed by the other studies
\citep{sadler02,best05,mauch07}, and b) the 2\sqdegs\ COSMOS field
samples a significantly smaller comoving volume at these low redshifts
compared to the almost all-sky surveys used by the other studies. This
verifies that both our selection method, as well as the derivation of
the LF are correct. Within the errors our lowest-redshift LF seems to
be best represented by the local LF derived by \citet[Sad02
  hereafter]{sadler02}, and we adopt this local LF for further
analysis of the evolution of our AGN population.

\begin{figure*}
\center{
\includegraphics[bb = 100 330 450 782]{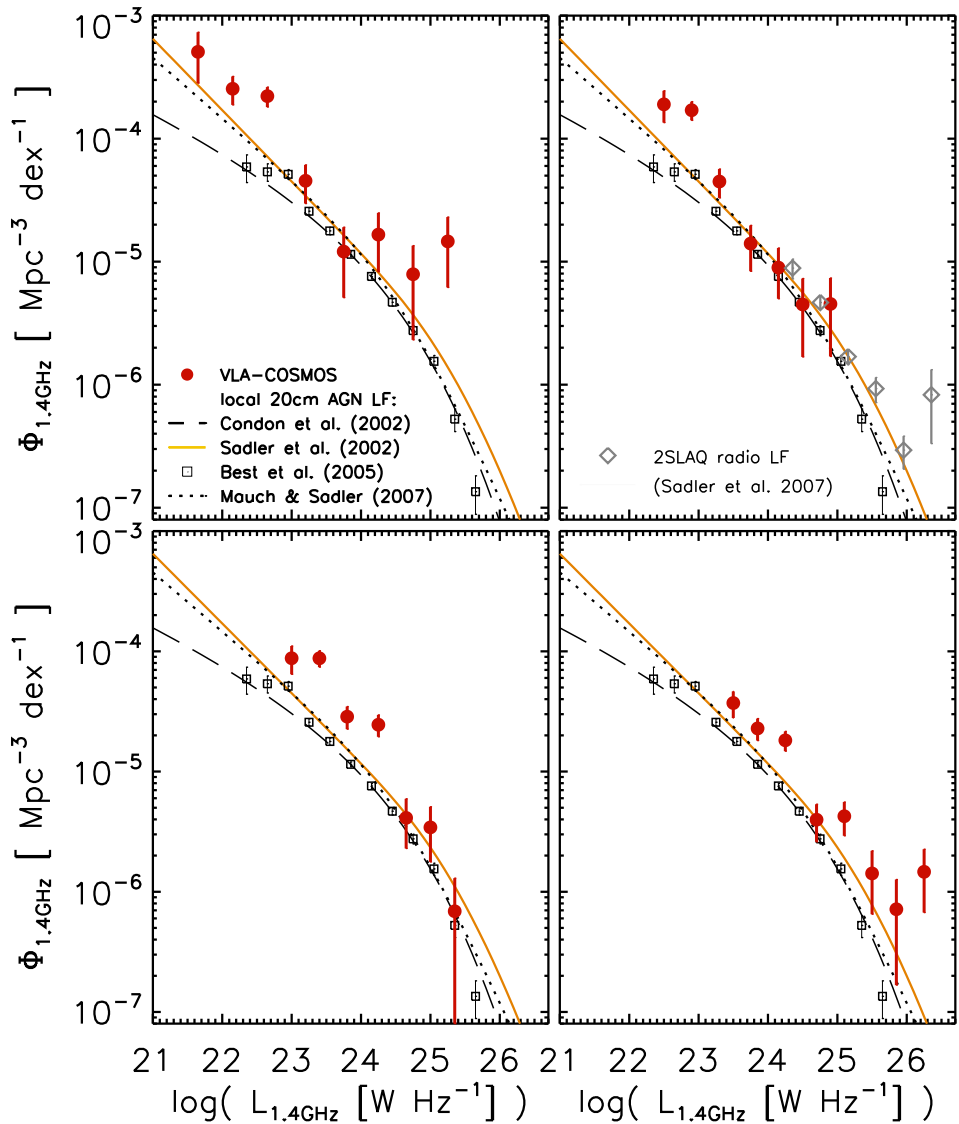} % for figure*
\caption{ 1.4~GHz luminosity functions (LFs) for AGN in the VLA-COSMOS
  survey (filled red circles), shown for four redshift ranges
  indicated in each panel.  The local AGN volume densities derived by
  \citet[dashed black line]{condon02}, \citet[solid orange
    line]{sadler02}, \citet[empty squares]{best05}, and \citet[dotted
    line]{mauch07} are also plotted in each panel. The empty symbols
  in the top right panel are the radio AGN volume densities derived
  for the redshift range $0.4<z<0.7$ by \citet{sadler07} for the 2SLAQ
  Luminous Red Galaxy survey sample, and scaled here to match our mean
  redshift value of 0.475 (see text for details).
  \label{fig:lf}}
}
\vspace{5mm}
\end{figure*}

In the top right panel in \f{fig:lf} \ we compare our derived AGN LF
($0.35<z\leq0.6$) with the volume densities computed by
\citet{sadler07} for a redshift range $0.4<z<0.7$ based on
2SLAQ
%\footnote{2dF-SDSS Luminous Red Galaxy and Quasi Stellar Object
%survey} 
Luminous Red Galaxy Survey and NVSS, FIRST data. We convert
their data to a base of $d\log{L}$, and scale them to match the mean
redshift of our redshift bin ($z=0.475$ compared to their $z=0.55$)
using pure luminosity evolution as obtained by \citet[note that this
  only decreases their radio luminosities by 10\%]{sadler07}. The LFs
are in excellent agreement. It is noteworthy that we well constrain
the volume densities at the low luminosity end ($\lesssim10^{25}$~\wh
), while the sample of \citet{sadler07} extends further out to
$1.6\times10^{27}$~\wh \ (see also \f{fig:lffull} , and
\s{sec:discussion} ).

The VLA-COSMOS AGN volume densities in the two highest redshift bins
($\rm 0.6<z\leq0.9$ and $0.9<z\leq1.3$) are shown in the bottom panels
in \f{fig:lf} .  The radio AGN LF at these redshifts has been studied
in full detail only for higher power AGN \citep[\lum$>10^{25}$~\wh ;
][]{dp90, willott01, waddington01}, while constraining the
low-luminosity end (sampled here) has been hampered by low number
statistics due to small field sizes.  \citet{cowie04} combined two
deep 1.4~GHz radio surveys of the HDF-N
%\footnote{Hubble Deep Field North} 
($40'$ in diameter; $5\sigma\sim40~\mathrm{\mu}$Jy in the
central region) and SSA13\
%footnote{SSA13} 
($34'$ in diameter;
$5\sigma\sim25~\mathrm{\mu}$Jy in the center) fields, and derived the
AGN radio LF in two separate redshift ranges similar to ours, and at
comparable luminosities to our work.  Our results are in qualitative
agreement with those derived by \citet[c.f.\ their
  Fig.~3]{cowie04}. However, their sample is a factor of $\sim7$
smaller than the one used here.

\section{ The evolution of radio AGN  }
\label{sec:lfevolv}

In this section we constrain the evolution of our low-power radio AGN
using the VLA-COSMOS AGN data (\s{sec:evolv} ). We further extend this
to high-power radio AGN using the model obtained by \citet{willott01}
based on high-power AGN samples drawn from the 3CRR, 6CE, and 7CRS
surveys (\s{sec:highpevolution} ).

\subsection{ The evolution of low-power radio AGN galaxies in the COSMOS field}
\label{sec:evolv}

The evolution of an astrophysical population is usually parameterized
by monotonic density and luminosity evolution of its local
luminosity function:

\begin{equation}
\label{eq:lfevolv}
\Phi_z(L) = (1+z)^{\alpha_D} \times \Phi_{z=0} \left[ \frac{L}{(1+z)^{\alpha_L}}
\right] 
\end{equation}

where $\alpha_D $ and $\alpha_L$ are the characteristic density and
luminosity evolution parameters, respectively, $L$ is the luminosity,
$\Phi_z(L)$ is the luminosity function at redshift $z$, and
$\Phi_{z=0}$ is the local luminosity function.

It is well known that strong degeneracy between luminosity and density
evolution exists, even if the observational data sample the turn-over
(`knee') of the LF at different cosmological times (see
e.g.\ \citealt{lefloc'h05}).  The VLA-COSMOS AGN sample in particular
lacks high luminosity objects that could constrain the turn-over of
the LF in all our redshift bins (see \f{fig:lf} ). Hence, we do not
attempt to try to break the luminosity -- density evolution
degeneracy, but we separately constrain both pure density (PDE;
$\alpha_L = 0$) and pure luminosity (PLE; $\alpha_D = 0$) evolution
based on our data.  We adopt the Sad02 local AGN LF ($\Phi_{z=0}$) as
the representative LF in the local universe, which is given by the
following analytical form:

\begin{equation}
\label{eq:sadler}
\Phi(L) = \Phi^* \left[ \frac{L}{L_*} \right]^{1-\alpha}
   \exp{ \left\{ -\frac{1}{2\sigma^2} \left[ \log{ ( 1+\frac{L} {L_*} )}
   \right]^2 \right\} } 
\end{equation}

where $\alpha = 1.58$, $\sigma=1.0$, $\Phi^* =
7.6\times10^{-6}$~Mpc$^{-3}$, and $L^*=2.1\times10^{24}$~W~Hz$^{-1}$
for their AGN population (scaled to the cosmology used here, and to
the base of $d \log L$).

For both PDE and PLE we derive the evolution by summing the $\chi^2$
distributions obtained for a large range of evolution parameters
$\alpha_L$ and $\alpha_D$ [($\alpha_L$,0) for PLE; (0,$\alpha_D$) for
PDE] in each particular redshift bin (excluding our first -- local --
redshift bin).  The uncertainty in $\alpha$ is then taken to be the
$1\sigma$ error obtained from the $\chi^2$ statistics. Our results
yield a pure density evolution with $\alpha_D=1.1\pm0.1$, or
alternatively a pure luminosity evolution with $\alpha_L=0.8\pm0.1$.

\begin{figure}[h!]
\includegraphics[bb =  54 320 486 792, width=\columnwidth ]{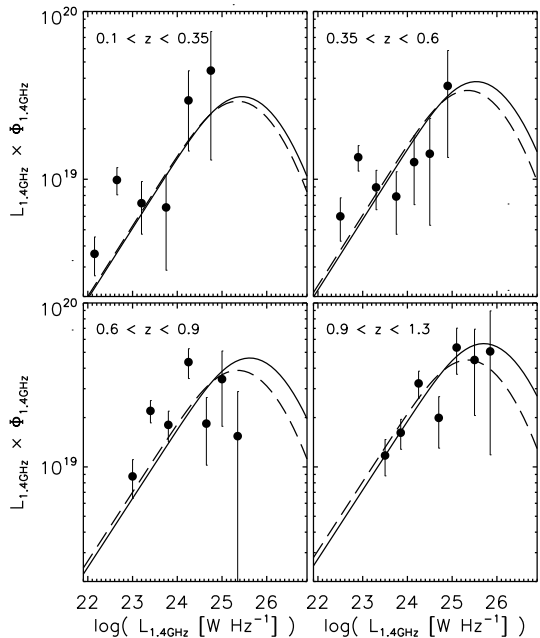}
\caption{ Luminosity density for the VLA-COSMOS AGN in four redshift
  ranges (filled circles). The  lines represent the best fit pure
  density (dashed line) and pure luminosity (solid line) evolution of
  the local 20~cm AGN luminosity function (see text for details).
  \label{fig:ld}}
\end{figure}

In \f{fig:ld} \ we show the luminosity density for our AGN in the four
redshift ranges defined in \f{fig:lf} . We also plot the luminosity
density given the derived PD and PL evolutions (lines). From the
figure it becomes obvious that we cannot distinguish between these two
types of evolution given our data, as our sample, tracing only the
lower luminosity end of the luminosity function,  appears to be
described equally well by both.  Thus, in the further analysis we will
take the range of the two best fit evolution models (taking also their
errors into account) as representative of the range of uncertainties.

In summary, our results imply either a pure luminosity evolution with
$L_*\propto (1+z)^{0.8\pm0.1}$ or a pure density evolution where
$\Phi_*\propto (1+z)^{1.1\pm0.1}$.  Regardless of which model is
physically more appropriate to describe the real cosmic evolution of
the VLA-COSMOS AGN population, both imply a modest evolution of radio
luminous AGN in the luminosity range of $\sim10^{22-25}$~\wh .

Our results are consistent with previous findings.  Based on a
$V/V_\mathrm{max}$ analysis out to a redshift of $\sim1$,
\citet{clewley04} have found no strong evolution of low radio
luminosity sources ($L_\mathrm{325MHz}\lesssim10^{25}$~\whs ,
corresponding to \lum~$\lesssim4.5\times10^{25}$~\wh ) at least up to
$z\sim0.6$. Further, \citet{sadler07} have constrained the evolution
of the luminosity function for radio luminous AGN
(\lum~$\sim10^{24-27}$~\wh ) out to $z\sim0.7$. Using the local AGN LF
given by \citet{mauch07} they have found a pure luminosity evolution
with $L_*\propto(1+z)^{2.0\pm0.3}$. The VLA-COSMOS AGN LF in the
redshift range $0.35<z\leq0.6$, and the \citet{sadler07} AGN LF
($0.4<z<0.7$) agree remarkably well (see top right panel in \f{fig:lf}
\ \& \f{fig:lffull} ).  If we parameterize the evolution of the
VLA-COSMOS AGN in the same way as described in \citet[using the same
  local LF]{sadler07}, we derive a pure luminosity evolution of
$1.2\pm0.4$ for the $0.35<z\leq0.6$ redshift bin.  It has to be
  noted, however, that the VLA-COSMOS and the 2SLAQ samples constrain
  different radio luminosity ranges (see \f{fig:lffull} ). Our
  results agree within the uncertainties with those of
  \citet{mauch07}, however they on average imply a slower
  evolution. This is consistent with a more modest evolution of low
  radio-power, compared to high radio-power, radio AGN
  \citep{willott01, waddington01, clewley04}. Our findings are also
in very good agreement with the recent results based on NVSS/FIRST and
the MegaZ-Luminous Red Galaxy catalog drawn from the SDSS
\citep{donoso08}.

\begin{figure*}%[h!]
\begin{center}
\includegraphics[bb = 100 330 450 782]{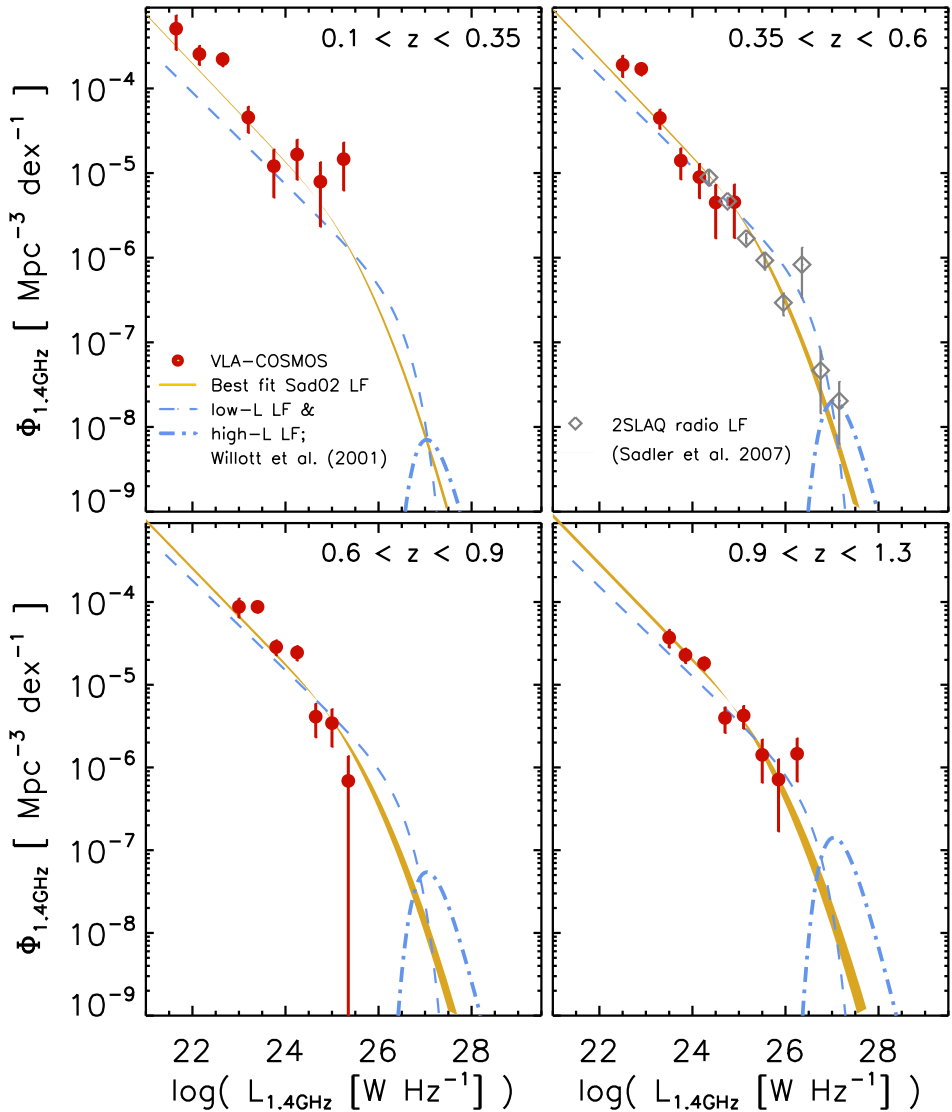}
\end{center}
\caption{ VLA-COSMOS AGN volume densities at 20~cm in four redshift
  ranges (red filled circles; analogous to \f{fig:lf} ). Also shown in
  each panel is our best fit evolution corresponding to the range
  given by pure density and pure luminosity evolution of the Sad02
  local LF and their corresponding errors (orange shaded curve), as
  well as the radio AGN luminosity function model given by
  \citet[model `C']{willott01} for their less luminous population
  (dashed blue curves), and high-luminosity population (dash-dotted
  blue curves; scaled to current cosmology and 1.4~GHz radio
  frequency; see text for details). In the top right panel the volume
  densities derived by \citet[scaled to match our mean redshift value
  of 0.475; see also \f{fig:lf} ]{sadler07} are shown.
  \label{fig:lffull}}
\end{figure*}

\subsection{ The evolution of high-power radio AGN }
\label{sec:highpevolution}

In order to quantify the contribution of different populations to the
comoving radio energy density at different cosmic times, {\em all}
radio AGN populations need to be considered. As low- and high- power
radio AGN seem to evolve in a different manner (see
\s{sec:introduction} ) and our VLA-COSMOS AGN sample only the low
power radio AGN, we need to make the following assumptions about the
evolution of the high power radio sources.

Based on the 3CRR, 6CE, and 7CRS radio surveys combined with complete
optical spectroscopy, \citet{willott01} have successfully modeled the
radio AGN LF using two radio populations -- a less powerful
(\lum~$\sim2.5\times10^{25-27}$~\wh ) population comprising both FR~I
and FR~II sources, and a powerful population
(\lum~$\gtrsim2.5\times10^{26}$~\wh ) comprising mostly FR~II sources.
They have modeled the evolution of the first population as a pure
density evolution up to a maximum redshift ($\sim0.7$) beyond which
any evolution ceases.  The evolution of the powerful population has
been assumed to change in density following a Gaussian distribution in
redshift, which was allowed to have a different shape beyond its
redshift peak at $z\sim2$ (see Tab.~1 in \citealt{willott01}). It is
worth noting that the Willott et al. model agrees well with the
\citet{dp90} steep spectrum model.

The VLA-COSMOS AGN sample constrains the faint end of the radio AGN
luminosity function, and here we use the \citet{willott01} model to
extend our radio LF to high powers.  In \f{fig:lffull} \ we compare
our LFs with the \citet{willott01} model, after the latter has been
converted to the current cosmology and the 151~MHz radio luminosities
scaled to 1.4~GHz.  The VLA-COSMOS AGN data and the Sad02 LF constrain
the radio LF more robustly at the faint end compared to the
\citet{willott01} model for their {\em less-powerful radio AGN}.
Thus, in the further analysis we will constrain the low-power radio
AGN LF and its evolution using the VLA-COSMOS sample as described in
the previous section, and we will use {\em only} the powerful radio
AGN model by \citet{willott01} to describe the evolution of high
radio-power AGN (see dash-dotted curves in \f{fig:lffull} ).

\subsection{ The evolution of the  comoving radio luminosity density for AGN
  galaxies } 
\label{sec:ldevolv}

At a specific cosmic time the integrated comoving luminosity density
represents the total power per unit comoving volume of a given
astronomical population.  Thus, if divided into distinct populations
of objects it traces their relative contribution to the overall power
output at a given redshift.  To estimate the contribution of low and
high radio power AGN to the AGN radio luminosity output at different
cosmic times, we investigate in the following the evolution of the
comoving 1.4~GHz luminosity density, $\Omega_\mathrm{1.4GHz}$, for our
VLA-COSMOS AGN (\lum~$\lesssim10^{25}$~\wh ) and the high-power AGN
(\lum~$\gtrsim10^{26}$~\wh ), adopting the \citet{willott01} model.

For a given redshift $\Omega_\mathrm{1.4GHz}$ has been computed by
integrating the luminosity density over the entire range of radio
luminosities. For the VLA-COSMOS AGN the luminosity density has been
constrained using our best fit PLE and PDE models (see curves in
\f{fig:ld} \ and \s{sec:evolv} ), and for the high-power population
using the \citet{willott01} model `C', with the corresponding errors
(see tab.~1 in \citealt{willott01} and \s{sec:highpevolution} ).

\begin{figure}[h!]
\includegraphics[bb =  54 360 486 792, width=\columnwidth ]{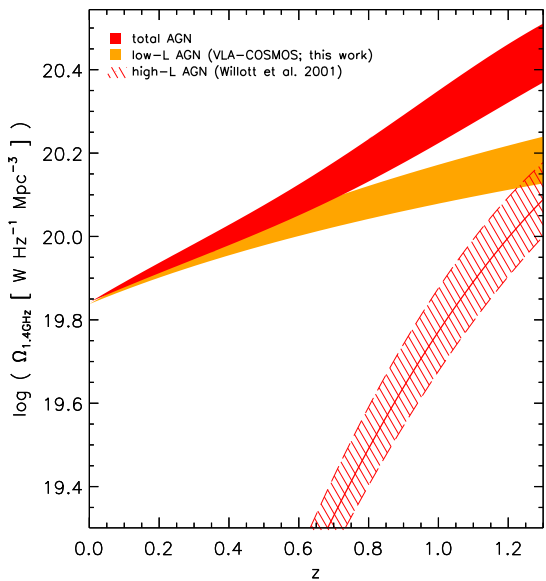}
\caption{ Evolution of the comoving 20~cm integrated luminosity density for
  VLA-COSMOS AGN (orange curve) galaxies since $z=1.3$. Shown is also the
  evolution of the high-luminosity radio AGN, adopted from \citet[hatched
  region; the thick and dashed lines correspond to the mean, maximum and
  minimum results, respectively]{willott01}.  The evolution for the total AGN
  population, obtained by co-adding the VLA-COSMOS and high-luminosity AGN
  energy densities, is shown as the red-shaded curve (see text for details).
  \label{fig:ldinteg}}
\end{figure}

In \f{fig:ldinteg} \ we show the evolution of the comoving 20~cm
integrated luminosity density for all radio AGN as well as the low and
high radio power AGN separately. The evolution of these two
populations is very different; high-power AGN evolve significantly
stronger than low-power AGN.  The total $\Omega_\mathrm{1.4GHz}$ for
radio AGN is dominated by low-power AGN at low redshifts
($z\lesssim0.7$) where the contribution of high-power radio sources is
negligible. However, at $z\gtrsim0.7$ high-power sources begin to
contribute significantly to the overall integrated luminosity density,
and at a redshift of $\sim1.3$ their contribution to the total AGN
$\Omega_\mathrm{1.4GHz}$ is comparable to the one of low-power
AGN. The implications of these populations for galaxy formation and
evolution, also out to higher redshifts ($z=2.5$), are discussed in
\s{sec:discussion} .

\section{ Properties of radio  AGN}
\label{sec:props}

In this section we outline and compare the properties of low and high
radio-power AGN in the local (\s{sec:propsLoc} ) and
intermediate-redshift (\s{sec:propsInt} ) universe. We find that
already by $z\sim1$ the host galaxies of low-power, VLA-COSMOS AGN
have built-up stellar and black hole masses comparable to the highest
mass galaxies observed locally. As their black hole masses are already
significant at these intermediate redshifts, this implies that they
will no longer be able to have a phase of high accretion (i.e.\
vigorous BH growth), consistent with numerous previous studies of
similar samples (e.g.\ \citealt{allen06, evans06, hardcastle06,
hardcastle07}).  For the high radio power AGN on the other hand,
extensive evidence exists in the literature that they accrete at high
rates -- representing a mode of substantial BH growth.

\subsection{Local universe}
\label{sec:propsLoc}

Various correlations are found in the literature between the presence
of emission lines in AGN and their e.g.\ radio power, black hole and
stellar mass, as well as environment and the galaxies' gas content. We
outline these below.

First, almost all FR~I -- low power -- radio galaxies are LERAGN,
while optical hosts of FR~IIs, which are typically more powerful
than FR~Is \citep{fr74,owen93,ledlow96}, usually have strong emission
lines\footnote{Note, however, that the correspondence between the FR
class and the presence of emission lines is not exactly one-to-one.}.
Recently, based on a large statistically significant sample of local
radio - optical sources (SDSS-NVSS-FIRST) \citet{kauffmann08} have
found that the fraction of radio AGN whose optical hosts have emission
lines in their spectra (predominantly HERAGN) is a strong function of
radio luminosity. This emission-line galaxy fraction is roughly
constant ($\sim40\%$) up to \lum~$\sim10^{25}$~\wh , beyond which it
steeply rises approaching $\sim80\%$ at $\sim4\times10^{25}$~\wh \
(see Fig.~3 in \citealt{kauffmann08}). This `critical' luminosity,
observed by Kauffmann et al., is remarkably close to the FR~I -- II
break luminosity, as well as to the power which roughly separates the
radio sources which show strong cosmological evolution from those
which do not (see e.g.\ \f{fig:ldinteg} \ and \s{sec:lfevolv}
). Furthermore, based on the results of Kauffmann et al., {\em the
luminosity of \lum~$\sim10^{25}$~\wh \ can be thought of as a rough
threshold between high- and low- excitation radio AGN}. Thus, most of
the low radio power AGN (\lum~$\lesssim 10^{25}$~\wh ) are LERAGN,
while the majority of powerful radio AGN (\lum~$\gtrsim 10^{25}$~\wh )
are HERAGN.

On the other hand, the fraction of emission line
(i.e.\ high-excitation) radio AGN in the local universe strongly {\em
  decreases} as a function of both stellar mass and velocity
dispersion (see Fig.~3 in \citealt{kauffmann08}). This implies that,
at least locally, high-excitation (or alternatively high radio power)
AGN tend to have lower stellar masses, as well as lower black hole
masses compared to LERAGN. Even further, the latter constitute the
most massive galaxies in the universe ($M_\odot> 10^{11}$~\msol ) that
preferentially occupy the centers of high galaxy density regions (Baum
et al.\ 1992, Best et al.\ 2005).

Furthermore, various studies in the literature have shown that HERAGN,
contrary to LERAGN, tend to show unusually blue off-nuclear continuum
colors \citep{baum92}, evidence of recent star formation
\citep{baldi08}, and often have distorted optical morphologies
suggesting that they have undergone a recent major merger
\citep{heckman86, baum92, baldi08}.
CO and HI observations of radio galaxies suggest larger amounts of
cold gas in powerful, compared to low-power, radio sources (see
e.g.\ Fig.~8 in Evans et al.\ 2005, see also Emonts et al.\ 2008), and
there is evidence supporting quantitatively larger amounts of dust,
and therefore gas \citep{leon01,sol05} in HERAGN compared to LERAGN (de
Koff et al.\ 2000, M\"{u}ller et al.\ 2004). A summary of the
properties of radio AGN in the nearby universe is given in
\t{tab:RAGNprops} .

\begin{deluxetable*}{lllll}
%\rotate
\tablecaption{Properties of LERAGN and HERAGN in the local
  universe\label{tab:RAGNprops}} 
\tablewidth{0pt} 
\tablehead{
   \colhead{} & 
   \multicolumn{2}{c}{LERAGN} & \multicolumn{2}{c}{HERAGN}
  \\ 
  \colhead{Property} & \colhead{} & \colhead{Ref.}  & \colhead{} & \colhead{Ref.}
}  
\startdata 
Object class & mainly FR~I & (1),(2) & mainly FR~II & (1),(2)\\
radio luminosity & \lum~$\lesssim10^{25}$~\wh  & (3) &
\lum~$\gtrsim10^{25}$~\wh & (3)\\ 
density environment & moderate-to-high  & (4),(5)& moderate-to-low & (4) \\ 
optical morphology & regular & (8),(9) & often distorted& (8),(9),(10)\\ 
optical color & red &(9),(11)& bluer (compared to LERAGN)& (10),(9)
\\ stellar mass & highest ($\gtrsim5\times10^{10}$~\msol) &(3),(11)& lower (compared to LERAGN)& (3),(12)\\ 
BH mass & highest ($\sim10^{9}$~\msol) &(9),(13)& lower (compared to LERAGN)& (3),(14)\\ 
ISM content & low &(15),(16),(17)& higher (compared to LERAGN)& (15),(16),(17),(18),(19)\\  
accretion mode & radiatively inefficient &(1),(2),(14),(20) & radiatively efficient & (1),(2),(20),(21),(22)
\enddata 
\tablecomments{ 
References: 
(1) \citealt{evans06}, 
(2) \citealt{hardcastle06}, 
(3) \citealt{kauffmann08}, 
(4) \citealt{baum92},
(5) \citealt{best05}, 
(8) \citealt{heckman86}, 
(9) \citealt{baum92},
(10) \citealt{baldi08},
(11) \citealt{smo08a},
(12) \citealt{tasse08},
(13) this work,
(14) \citealt{chiaberge05},
(15) \citealt{leon01},
(16) \citealt{dekoff96},
(17) \citealt{muller04},
(18) \citealt{evans06},
(19) \citealt{emonts08},
(20) \citealt{hardcastle07}, 
(21) Barthel 1989, 
(22) Haas 2004
}
\end{deluxetable*}

\subsection{Intermediate redshift}
\label{sec:propsInt}

In this section we investigate the properties of low and high
radio-power AGN at intermediate redshifts, and compare these to the
properties of their local counterparts.  In \s{sec:lf} \ and
\s{sec:lfevolv} \ we have derived the 1.4~GHz radio luminosity
function, and its evolution, for a sample of $\sim600$ radio luminous
AGN ($0.1<z<1.3$) drawn from the VLA-COSMOS survey. One of the main
characteristics of this radio AGN sample is that it consists of
predominantly ($96\%$) low-power AGN with \lum~$\lesssim10^{25}$~\wh
. \citet{ledlow96} have shown that the threshold between weak FR~I
types of galaxies and powerful FR~IIs is a function of the host
galaxy's optical ($R$ band) magnitude, $\mathrm{M_R}$. Hence, to
better constrain the properties of the VLA-COSMOS AGN in
\f{fig:mrlrad} \ we plot them in the \lum \ vs.\ $\mathrm{M_R}$
plane. Almost all ($98\%$) of our AGN occupy the weak radio AGN, FR~I
region in this plane.

\begin{figure}[h!]
\includegraphics[bb =  54 370 486 792, width=\columnwidth ]{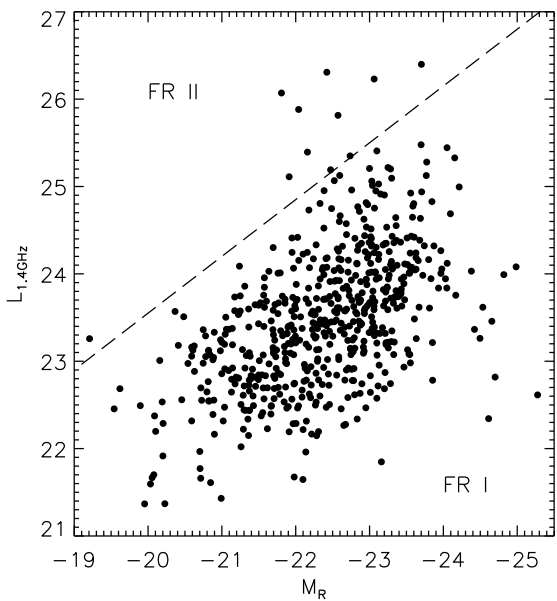}
\caption{ Monochromatic 1.4~GHz radio power for VLA-COSMOS AGN as a
  function of their host galaxy absolute $R$ band magnitude. The
  dashed line corresponds to the separation between FR~I and FR~II
  types of galaxies given by \citet{ledlow96}. Note that almost all of
  the VLA-COSMOS AGN occupy the low-power FR~I region of this
  plane. 
  \label{fig:mrlrad}}
\end{figure}

\begin{figure*}[t]
\includegraphics[bb = 94 400 486 750, scale=0.7]{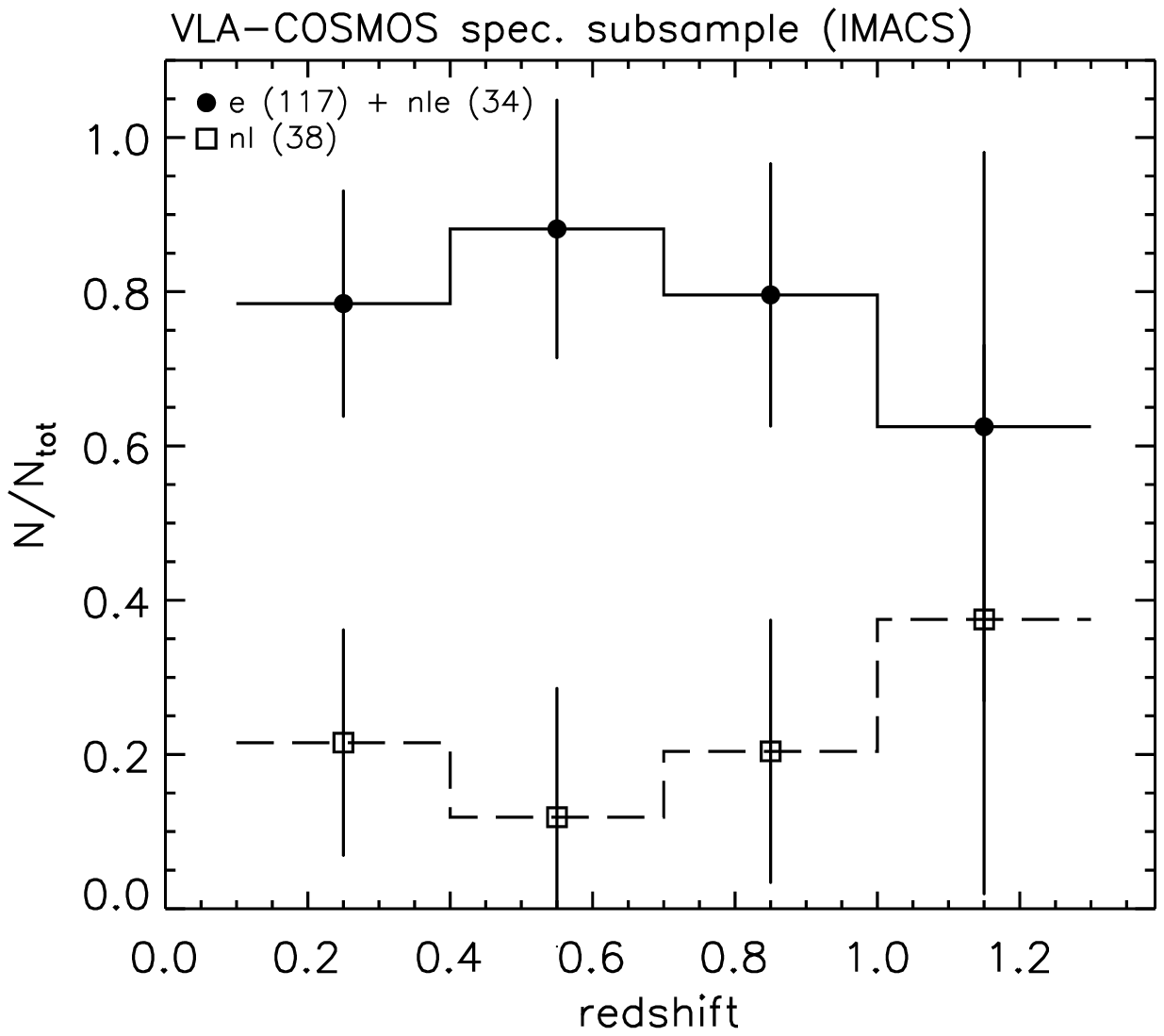}
\includegraphics[bb = 175 400 486 750, scale=0.7]{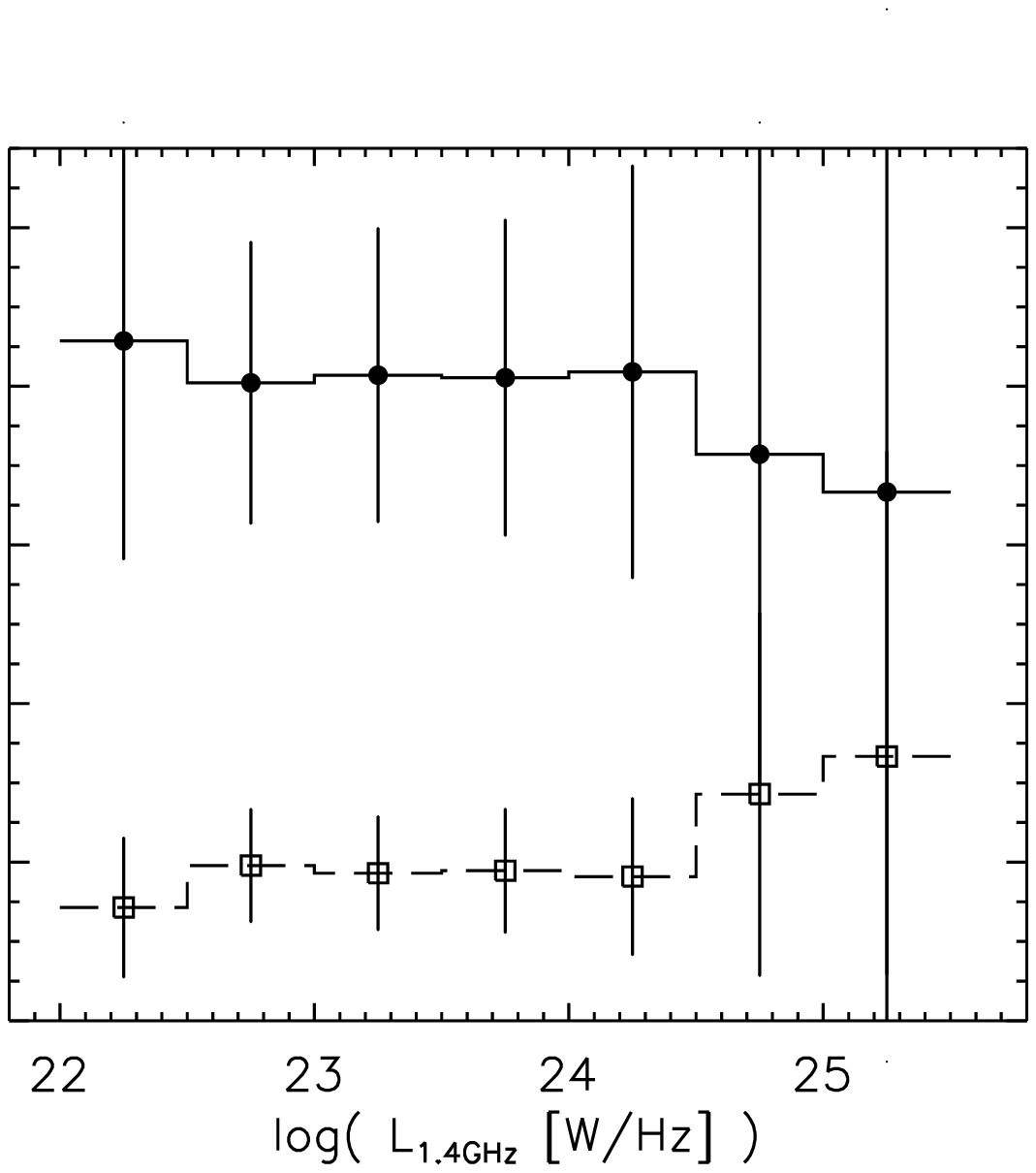}
\caption{ The fraction of VLA-COSMOS AGN with available
  Magellan/IMACS spectroscopy \citep{trump07} as a function of
  redshift (left panel) and 1.4~GHz radio luminosity (right
  panel). The galaxies have been divided into two sub-samples, i)
  narrow emission line (nl) galaxies, and ii) elliptical galaxies (e)
  plus composite objects showing a red galaxy continuum and narrow
  emission lines (nle; see \citealt{trump07} for details). These two
  subsamples roughly correspond to high- and low-excitation AGN,
  respectively (see text for details). The quantity and symbols for
  objects of each type is indicated in the left panel. Note that
  there is a slight (although noisy) trend showing that the fraction
  of narrow emission line objects (i.e.\ roughly high-excitation AGN)
  increases as a function of both redshift and radio luminosity.
  \label{fig:imacsfrac}}
\end{figure*}

In order to get a further insight into the composition of the
VLA-COSMOS AGN, we make use of the sources with available optical
spectroscopy.  \citet{trump07} have carried out a spectroscopic survey
of X-ray and radio selected AGN candidates in the COSMOS field (using
the Magellan/IMACS instrument). They have classified their sources
into a) broad emission line AGN (`bl'), b) narrow emission line AGN
(`nl'), c) red galaxies with only absorption features in their
spectrum (`e') and d) hybrid objects showing a mix of narrow emission
lines and a red galaxy continuum shape (`nle').  Combining a radio AGN
detection with this classification scheme, we can very roughly
consider the red and hybrid galaxy spectroscopic classification
(`e'$+$'nle') as a proxy for LERAGN, and the narrow emission line AGN
classification (`nl') as a proxy for HERAGN. About 35\% of our
VLA-COSMOS AGN have an available classification based on IMACS
spectroscopy, and in \f{fig:imacsfrac} \ the fractions of `e$+$nle'
and `nl' galaxies are shown as a function of both redshift (left
  panel) and radio luminosity (right panel). The absorption line and
hybrid galaxies dominate the VLA-COSMOS sample at a constant
$\sim80\%$ level at all redshifts ($z\leq1.3$), while the narrow
emission line AGN contribute about $\sim20\%$ at all redshifts. There
is an indication that in the highest redshift bin ($1<z<1.3$), where
our most luminous (\lum~$>10^{25}$~\wh ) radio AGN are observed (see
\f{fig:lf} ), the fraction of narrow emission line AGN rises. However,
given the large error-bars in this redshift range (due to the low
number of spectroscopically observed radio AGN) we cannot draw any
robust conclusions regarding this. In the right panel of
  \f{fig:imacsfrac} \ we show the fraction of `e$+$nle' and `nl'
  galaxies as a function of 1.4~GHz power. Although the number of
  sources with \lum~$\gtrsim10^{25}$~W~Hz$^{-1}$ is low in the
  VLA-COSMOS AGN sample, there is an indication that the fraction of
  narrow-line objects (thus roughly HERAGN) increases beyond
  $\sim10^{25}$~W~Hz$^{-1}$, consistent with the findings in the local
  universe \citep{kauffmann08}.

Assuming that the spectroscopic sub-sample represents well the full
sample (note that this is a rather robust assumption as shown in S08a;
see their Figs.~3~and~21), we conclude that LERAGN dominate the
VLA-COSMOS radio AGN sample.  Thus, similar to the findings in the
local universe, our low radio power AGN at intermediate redshift are
preferentially LERAGN.

\begin{figure}[h!]
\includegraphics[bb =   54 360 486 750, width=\columnwidth
]{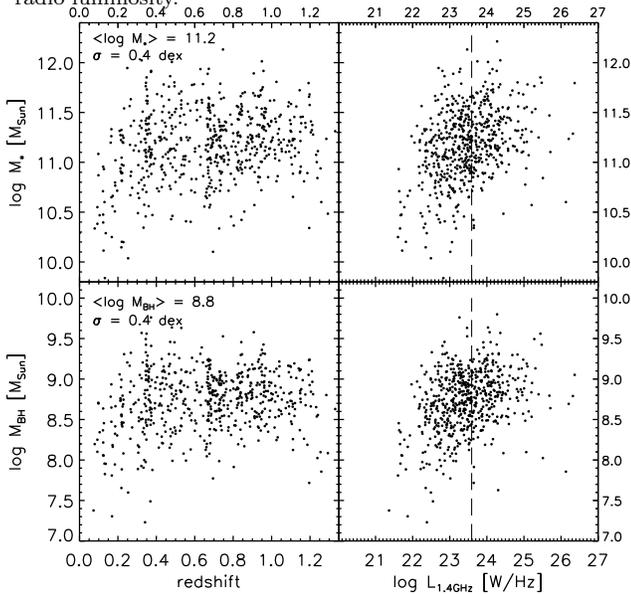}
\caption{ Logarithm of stellar (top panels) and black hole (bottom
  panels) masses for the VLA-COSMOS AGN as a function of redshift
  (left panels) and 1.4~GHz radio luminosity (right panels). The
  stellar masses were computed via SED fitting using the \citet{bc03}
  stellar population synthesis models (\citealt{chabrier03} IMF; see
  S08a for full details). The BH masses were derived using
  eq.~\ref{eq:bhmass}, and corrected for passive luminosity evolution
  (see text for details).  The dashed line in the right column
  corresponds to the lowest 1.4~GHz luminosity observable out to
  $z=1.3$ given the VLA-COSMOS Large Project detection limit of
  $\sim50~\mu$Jy~beam$^{-1}$. \label{fig:mass}}
\end{figure}

Stellar masses, using a \citet{chabrier03} initial mass function
(IMF), have been computed for the entire VLA-COSMOS galaxy sample
($z\leq1.3$) by S08a.  In \f{fig:mass} \ (top panels) we show the
stellar masses for our VLA-COSMOS AGN as a function of redshift and
1.4~GHz luminosity. The median stellar mass of our radio AGN is
$1.6\times10^{11}$~\msol , i.e.\ $<\log M_*> = 11.2$ (in any given
redshift range) with a $1\sigma$ scatter of 0.4~dex. This is
consistent with the stellar masses of the most massive, local galaxies
(e.g.\ \citealt{baldry04, best05}).  In addition, it is worth noting
that their average rest-frame colors are consistent with red galaxy
colors (see e.g.\ Fig.~9 in S08a).

We further compute the BH masses for the full VLA-COSMOS AGN sample
using the local correlation given by \citet{marconi03} which
relates the K-band luminosity to the BH mass:
\begin{equation}
\label{eq:bhmass}
\log_{10}M_\mathrm{BH} = 8.21 + 1.13 \times (\log_{10}L_K - 10.9)
\end{equation}
% where $M_\mathrm{BH}$ is the black hole mass (in solar units), and
$L_K$ the rest-frame K-band luminosity (also in solar units).  The
scatter in the relation is 0.5~dex \citep{marconi03}.  The K-band
rest-frame luminosity for our VLA-COSMOS AGN was computed via SED
fitting as described in detail in S08a.  In order to take into
  account passive luminosity evolution (see \citealt{hopkins06b} and
  references therein) we change the normalization constant of the
  above relation as a function of redshift following the results from
  \citet[][see short-dashed line in the bottom left panel in their
    Fig.~2]{hopkins06b}.  In \f{fig:mass} \ (bottom panels) we plot
  the estimated BH masses as a function of both redshift and radio
  1.4~GHz luminosity.  The median BH mass of our radio AGN is
  $\log_{10}M_\mathrm{BH} = 8.8$ with a standard deviation of 0.4.
Note that the apparent trend of BH mass with radio luminosity (bottom
right panel in \f{fig:mass} ) reflects the dependence of radio
luminosity on redshift in our radio flux limited sample (see Fig.~16
in S08a) rather than a real trend.

The black hole masses of our intermediate redshift radio AGN are
comparable to the highest black hole masses known in the local
universe (see e.g.\ \citealt{md02, md04}). This implies that
the low-power radio AGN have assembled their BH masses already by
these intermediate redshifts, and that their BHs cannot grow
significantly since $z\sim1.3$. Therefore, they must be in a mode of
modest growth in BH mass.  This result is in agreement with
numerous findings in the literature implying that LERAGN accrete
radiatively inefficiently, at sub-Eddington rates \citep{allen06,
  evans06, hardcastle06, hardcastle07}.

Given all of the above, the composition of our VLA-COSMOS AGN, the
majority of which are shown to be LERAGN, is consistent with the
properties of low radio-power AGN in the local universe: Our
intermediate redshift AGN 
have already by $z\sim1$ assembled both their stellar and black
hole masses, comparable to the high-mass end of the galaxies known
today.

On the other hand, powerful radio galaxies at high redshifts tend to
show strong emission lines in their optical spectra (Rawlings et
al.\ 1989, Baum \& Heckman 1989, Rawlings \& Saunders 1991; Willott et
al.\ 1999, 2000), and are often associated with ongoing star-formation
(Archibald et al.\ 2001, Greve et al.\ 2006, Seymour et al.\ 2008) as
suggested by their bluer rest-frame colors (compared to red-sequence
galaxies). Again, this is consistent with local observations of such
sources. In numerous studies in the literature
\citep[e.g.][]{barthel89, haas04} these objects have been shown to
accrete radiatively efficiently at high ($\sim$~Eddington) accretion
rates.  Thus these HERAGN present a mode of significant BH growth --
unlike the low radio power LERAGN.

To summarize, the properties of local and intermediate-redshift radio
AGN, as shown above, suggest that radio activity is triggered in
similar populations of objects independent of their redshift, i.e.
cosmic time.  This is in agreement with the finding that the
rest-frame optical colors of radio source host galaxies do not change
with redshift (\citealt{barger07}, S08a, \citealt{huynh08}). In
addition, there is converging evidence that low radio power AGN
reflect a modest -- radiatively inefficient -- mode of BH growth,
while high radio power AGN are undergoing a phase of significant --
radiatively efficient -- BH growth.

\section{The star formation quenching and radio-AGN triggering rates}
\label{sec:sfagntrigrates}

One of the aims of this work is to study the link between radio AGN
activity and star formation from an observational perspective. Thus,
in order to investigate whether the processes responsible for the
build up of the red galaxy sequence and the triggering of radio AGN
mode are related or not, in this section we study the stellar
  mass properties of the red parent (\s{sec:parent} ) and radio AGN
  samples (\s{sec:sm} ), and derive and compare the rates for the
star formation quenching (\s{sec:sfq} ) and (low radio power) AGN
triggering (\s{sec:dutycyc} ). We find no evidence that these two
processes are related. This however does not exclude the possibility
that the radio AGN phase may be responsible for preventing already
assembled massive galaxies to grow higher in mass, as explored in
\s{sec:feedevolv} .

\subsection{The parent galaxy sample}
\label{sec:parent}

In order to derive the rates for star formation quenching and
radio-AGN triggering we need to define a control-parent galaxy sample
of our VLA-COSMOS radio AGN. We define this sample using the COSMOS
photometric redshift catalog \citep{ilbert08} from which we select
galaxies with the same optical magnitude, redshift and color criteria
applied to our VLA-COSMOS AGN (see \s{sec:sample} ). The rest-frame
optical color P1 (constrained from the $3200-5800$~\AA\ range), which
was used to select our VLA-COSMOS AGN is not available for the
galaxies in the COSMOS photometric redshift catalog. Nonetheless, as
the NUV-NIR galaxy SED has been shown to be a one-parameter family
\citep[e.g.][]{obric06,smo06}, we can safely utilize another
rest-frame color for the selection of the parent sample. In
\f{fig:p1ub} \ we show the correlation between $U-B$ and P1 rest-frame
colors for all VLA-COSMOS galaxies ($z\leq1.3$), and derive an analytic
relation correlating P1 and $U-B$ (indicated in \f{fig:p1ub} ). The
criterion of P1~$>0.15$, used to select the VLA-COSMOS AGN sample,
corresponds to $U-B>0.71$. Thus we select the red galaxy parent sample
by requiring $I_{auto}\leq 24$, $0.1<z_{phot}\leq1.3$, and
$U-B>0.71$. The $U-B$ distribution for all (red and blue) galaxies,
that satisfy the magnitude and redshift criteria, is shown in
\f{fig:ub} . Note that the adopted $U-B$ threshold, based on the
comparison with the $P1$ values, corresponds almost exactly to the
$U-B$ value which separates the red from the blue galaxies. Our
selection yields 21,525 galaxies in the red galaxy control-parent
sample.

\begin{figure}[h!]
\includegraphics[bb = 54 360 486 750, width=\columnwidth
]{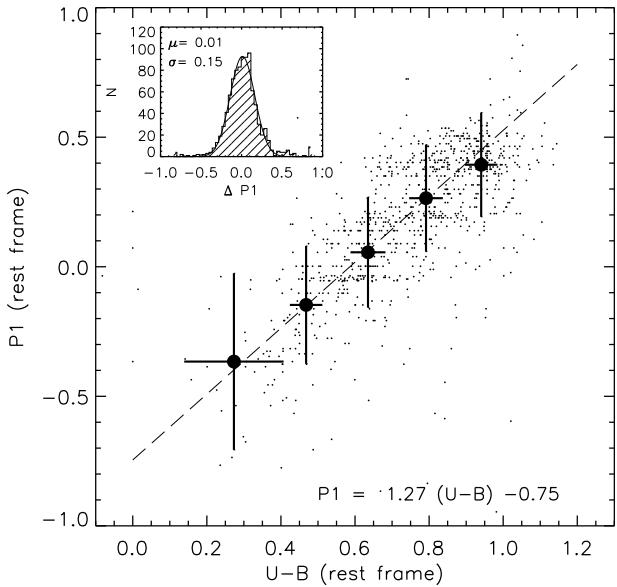}
\caption{ Comparison of the P1 and U$-$B rest frame colors for 941
  VLA-COSMOS radio--optical sources (defined in S08a; small dots). The
  dashed line shows the best linear fit (the analytic form is
  given in the bottom right of the panel), obtained using the
  median (P1,U$-$B) values (large dots). The inset shows the
  dispersion of the fitted relation, where for a given galaxy
  $\Delta$~P1 is the difference between its P1 value computed using
  its U$-$B rest-frame color (obtained by \citealt{ilbert08})
  and the one derived directly via SED fitting (as described in
  S08a). \label{fig:p1ub}}
\end{figure}

\begin{figure}[h!]
\includegraphics[bb = 54 450 486 730, width=\columnwidth
]{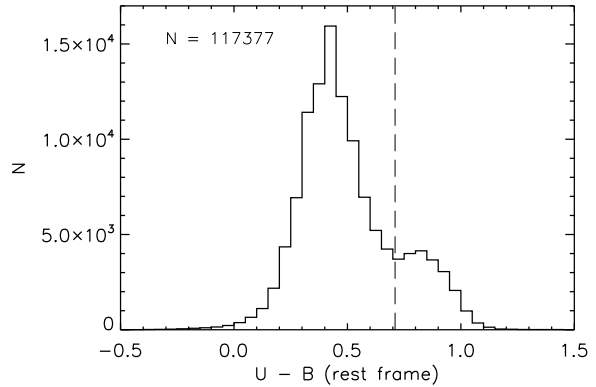}
\caption{ $U-B$ rest frame color distribution for 117,377 galaxies
  ($I_{auto}\leq24$; $0.1<z_{phot}\leq1.3$) drawn from the COSMOS
  photometric redshift catalog. The red galaxy parent sample of our
  VLA-COSMOS radio AGN is selected by applying $U-B>0.71$ (dashed
  line) to this distribution (this is equivalent to P1~$>0.15$ that
  was used to select our radio AGN; see \f{fig:p1ub}
  ). \label{fig:ub}}
\end{figure}

\subsection{The stellar mass properties of VLA-COSMOS AGN}
\label{sec:sm}

\begin{figure*}
\center{
\includegraphics[bb =  54 360 486 792]{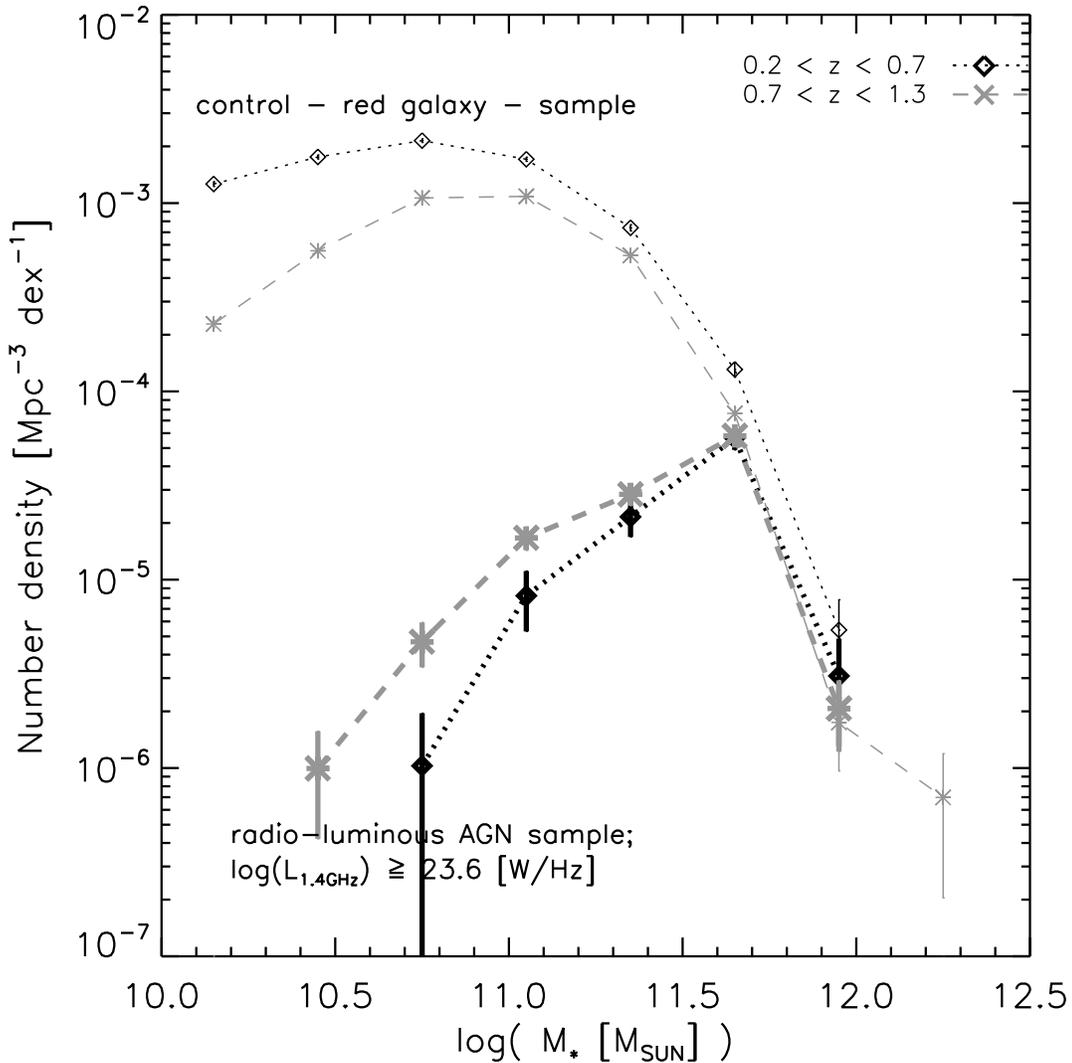} 
\caption{Stellar mass function, derived using the $1/$\Vm\ method, of
  the VLA-COSMOS radio AGN (volume limited,
  \lum~$\gtrsim4\cdot10^{23}$~\wh ; bold symbols) in two redshift
  ranges, indicated in the panel. For comparison, the stellar mass
  function of the red, parent galaxy sample in the same redshift
  ranges is also shown.
  \label{fig:mf}}
}
\vspace{5mm}
\end{figure*}
\begin{figure}
\center{
\includegraphics[bb = 54 360 486 792, width=\columnwidth]{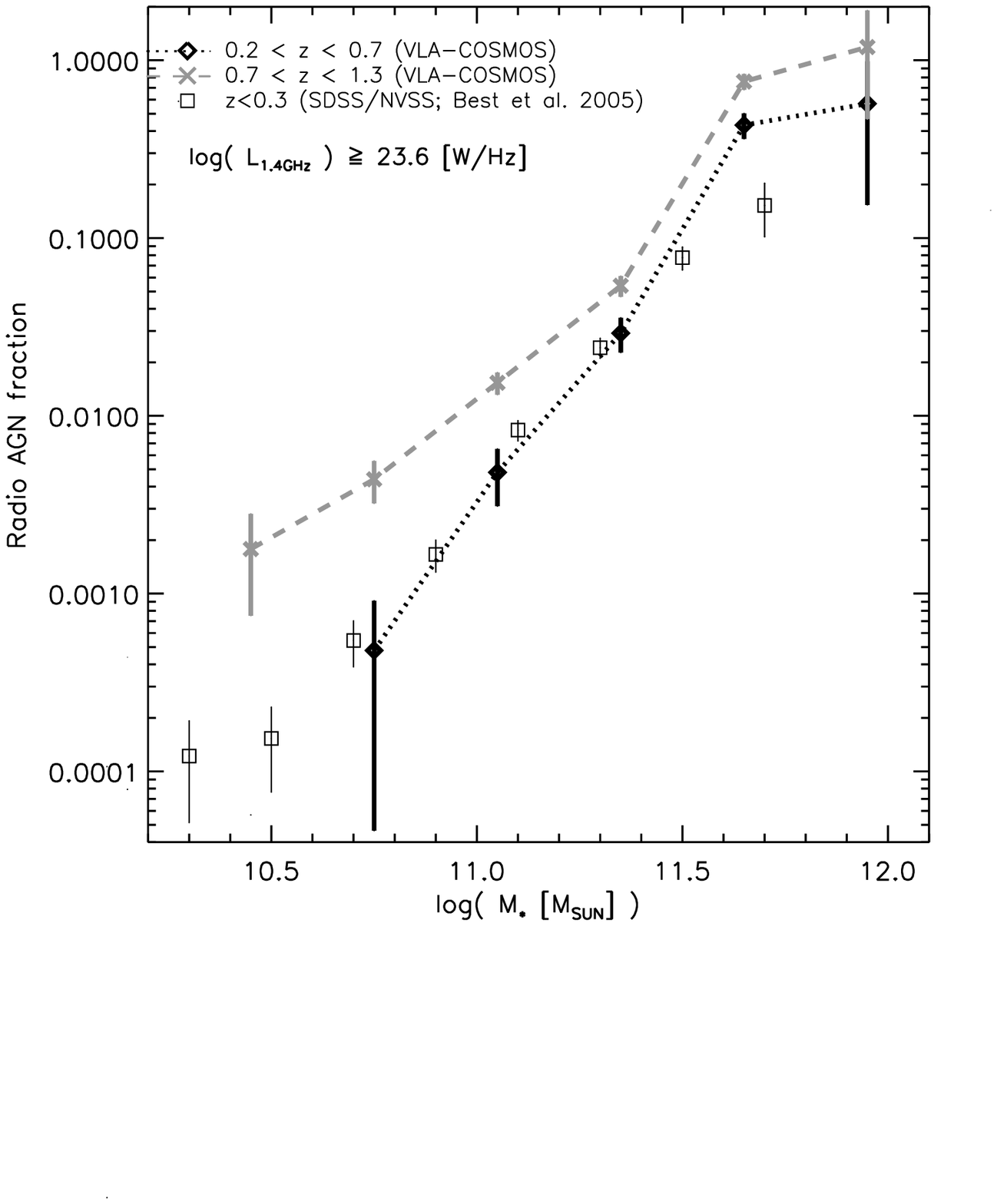}
\caption{ Fraction of radio luminous, VLA-COSMOS, AGN
  (\lum~$\gtrsim4\cdot10^{23}$~\wh ) as a function of stellar mass in
  two redshift ranges (bold symbols). Also shown is the radio AGN
  fraction based on the local ($z<0.3$) SDSS/NVSS samples with the
  same radio power limit imposed \citep{best05}. Note that no
  evolution of the radio AGN fraction is seen out to $z=0.7$.
  \label{fig:mfrac}}
}
\vspace{5mm}
\end{figure}

Using the $1/$\Vm\ method we derive the stellar mass function (SMF) in
two redshift bins ($0.2\leq z\leq0.7$ and $0.7<z\leq1.3$) for both our
VLA-COSMOS AGN (\lum $\gtrsim4\cdot10^{23}$~\wh\ which roughly
corresponds to a radio volume limited sample), and the red parent
galaxy sample.  The stellar mass functions are shown in \f{fig:mf}
. The SMF for the red galaxy sample agrees well with the recent
results based on the GOODS-MUSIC galaxy sample
\citep{fontana06}. Similarly to what has been shown in the local
universe for radio luminous AGN \citep{best05}, at intermediate
redshifts the SMFs of our radio AGN sample are strongly biased toward
high stellar masses. Interestingly, the redshift evolution of the SMFs
of the parent galaxy and radio AGN samples is reversed. While for a
given stellar mass the comoving number density of red galaxies
decreases with redshift, the number density of radio luminous AGN
increases. Our results are in good agreement with those of
\citet{tasse08} based on radio observations of the XMM-LSS field.

We compute the fraction of radio luminous AGN as the ratio of the
above derived mass functions in a given stellar mass bin.  In
\f{fig:mfrac} \ we show the radio AGN fraction as a function of
stellar mass in the two adopted redshift ranges, and compare it to the
results based on the local SDSS and NVSS surveys \citep{best05}. The
radio AGN fraction in our lower redshift ($0.2<z<0.7$) bin agrees
remarkably well with the local findings implying insignificant or
absent evolution of the radio AGN fraction out to $z\sim0.7$ at all
stellar masses. However, there is a significant change in the fraction
of radio luminous AGN at higher redshifts ($0.7<z<1.3$), in particular
for host galaxies with stellar masses lower than $\log{M_*} \sim
11.3$. We investigate this further in \s{sec:dutycyc} .

\subsection{The star formation quenching rate}
\label{sec:sfq}

Following Bundy et al.\ (2008), we define the star formation quenching
rate, $\dot{Q}$, as the fraction of all galaxies in a given stellar
mass bin that migrate to the red sequence per Gyr. This migration can
occur through processes such as mass build up (via `wet' mergers) or
gas consumption (fading of an already massive galaxy to red colors).
In \f{fig:alpha} \ (left panel) we plot the fraction of red galaxies
(as defined in \s{sec:parent} ) relative to the number of all galaxies
(see \f{fig:ub} ) as a function of cosmic time for various stellar
mass bins. The red galaxy fractions have been derived in the same way
as the radio AGN fraction (see \s{sec:sm} ), but using the red and
total galaxy samples.  The the change in radio fraction per Gyr
($\alpha$) of this red-galaxy-sequence build-up with cosmic time,
shown in the left panel of \f{fig:alpha} , is then equivalent to the
star formation quenching rate. We find that as stellar mass increases
from $\log{M_*}$ of 10.6 to 11.7~\msol\ $\dot{Q}$ decreases from
6~\%~Gyr$^{-1}$ to 0.6~\%~Gyr$^{-1}$. Note also that for the highest
stellar masses the red sequence is already almost fully in place by
$z\sim1$.

\begin{figure*}
\center{
\includegraphics[bb = 134 360 486 792, scale=0.61]{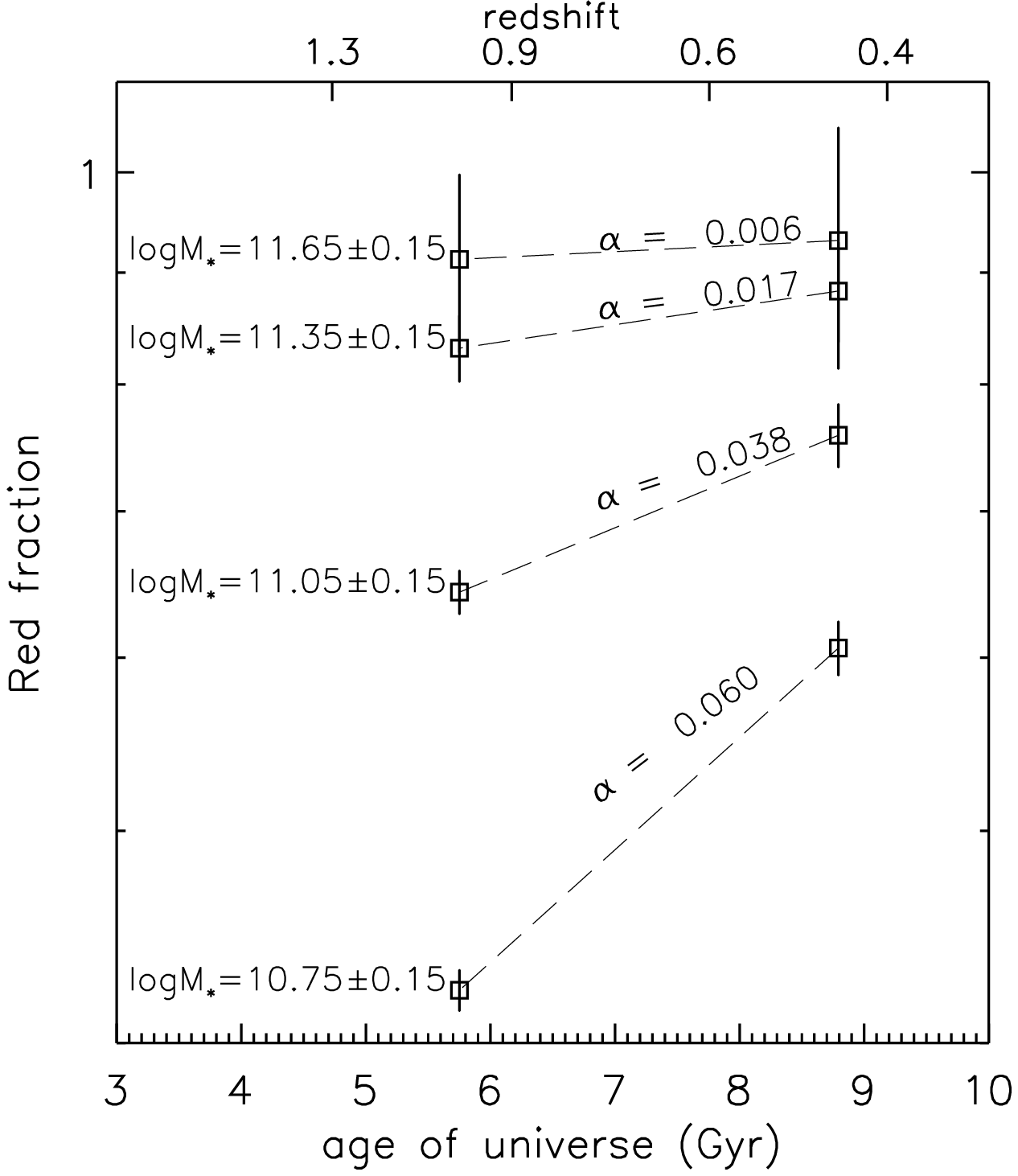} 
\includegraphics[bb= 84 360 486 792, scale=0.61]{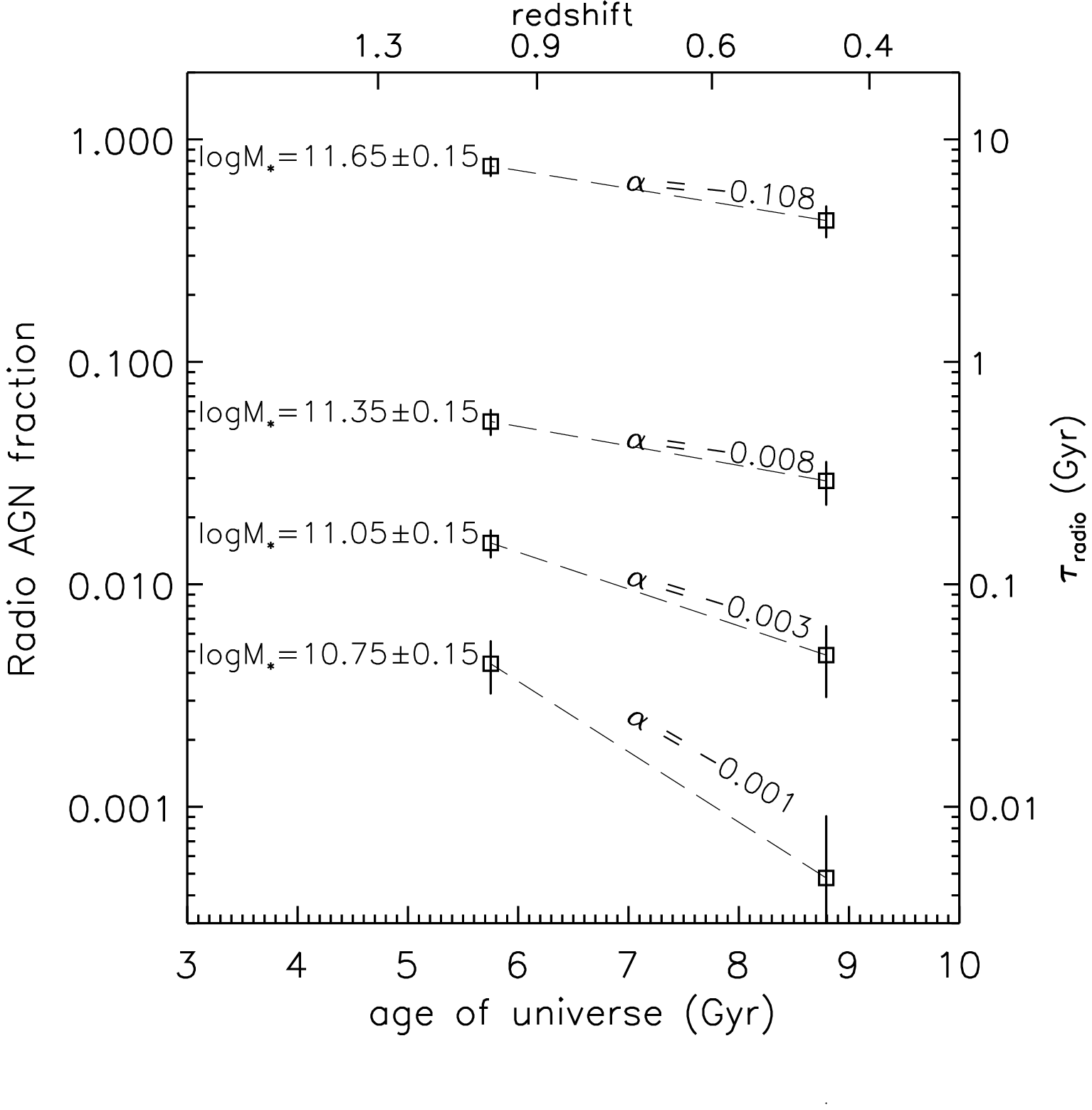}
\caption{ Left panel: The fraction of red galaxies ($P1>0.15$; equivalent to
  $U-B>0.71$; left panel), compared to all galaxies, as a function of
  cosmic age for different stellar mass ranges (indicated in the
  panel). The right panel shows the fraction of radio AGN in the red,
  parent galaxy sample (as defined in the text) as a function of
  cosmic age for the various stellar mass bins. The fractional change
  with cosmic time has been characterized with a linear slope ($\alpha$),
  which is indicated for each stellar mass bin. The radio AGN fraction
  can be related to the average time a galaxy spends as a radio AGN
  ($\tau_\mathrm{radio}$; see text for details), shown on the
  right-hand y-axis in the right panel. Note the different y scales in
  the left and right panels. Note also that there is no correspondence
  between the increase of the red galaxy fraction, and the decrease of
  the radio AGN fraction with cosmic time.
  \label{fig:alpha}}
}
\vspace{5mm}
\end{figure*}

\subsection{The  triggering rate of low radio-power AGN}
\label{sec:dutycyc}

In \s{sec:sfq} \ we have derived the star formation quenching rate
based on the build-up of the red galaxy population as a function of
cosmic age (left panel in \f{fig:alpha} ). In the right panel in
\f{fig:alpha} \ we now show the radio volume limited
(log~\lum~$\geq23.6$) fraction of our VLA-COSMOS AGN relative to the
abundance of the red parent galaxies as a function of cosmic time, for
the various stellar mass ranges as shown in \f{fig:mfrac} .  Contrary
to the red galaxy fractions, our results yield that the radio AGN
fraction decreases with cosmic age. At all stellar masses
($10.6<\log{M_*}<11.7$) a larger fraction of radio luminous AGN is
present at earlier, compared to later, cosmic times. Furthermore, the
fraction of radio luminous AGN decreases more steeply with cosmic time
for higher stellar masses. For example, at $\log{M_*}\sim11.65$ the
rate of decrease is 10.8~\%~Gyr$^{-1}$, while at $\log{M_*}\sim10.75$
it is only 0.1~\%~Gyr$^{-1}$.

Given the fraction of radio AGN at a given cosmic time, we can
estimate the average time that each single massive galaxy spends as a
radio AGN, $\tau_\mathrm{radio}$.  Assuming that massive galaxies,
that form the parent sample of our radio AGN, form around $z\sim2.5-3$
and survive to $z\sim0$, their lifetime is then $\sim10$~Gyr.
Multiplying this lifetime by the fraction of observed radio AGN then
yields an approximate estimate of $\tau_\mathrm{radio}$. At a given
cosmic time, $\tau_\mathrm{radio}$, which is in the range of roughly
$10$~Myr to $5$~Gyr for our VLA-COSMOS AGN (see right panel in
\f{fig:alpha} ), is thus related either to the average duration of a
single radio-episode or to the frequency of radio mode re-triggering.

Independent calculations of radio source lifetimes,
$t_\mathrm{radio}$, result in lifetimes of a few times 0.001~Gyr to a
few times 0.1 Gyr (Alexander \& Leahy 1987; Shabala et
al.\ 2008). Assuming that the radio source lifetime does not
significantly change with cosmic time, it is most likely that the
computed $\tau_\mathrm{radio}$ reflects the occurrence of multiple
radio-phases of a single massive galaxy at a certain cosmic time. A
radio `on' phase of 0.001-0.1 Gyr, combined with the derived
$\tau_\mathrm{radio}$, implies then that radio AGN activity has been
triggered more often in the past than it is today, and that at a given
cosmic time it is more often triggered in massive galaxies compared to
lower mass galaxies.

The {\em minimum} radio AGN triggering rate can be obtained by
assuming the maximum $t_\mathrm{radio}$ of $0.1$~Gyr, and taking into
account the above derived $\tau_\mathrm{radio}$ ($0.010-5$~Gyr). It is
then 0.1~\%~Gyr$^{-1}$ to 50~\%~Gyr$^{-1}$ for the lowest and highest
stellar mass bins, respectively. Yet, in the previous section we have
computed a star formation quenching rate of $6$~\%~Gyr$^{-1}$ for the
lowest and $0.6$~\%~Gyr$^{-1}$ for the highest stellar mass bins (see
left panel in \f{fig:alpha} ). Hence, comparing these estimates yields
that our derived radio AGN triggering rates are by many factors off
the derived star formation quenching rates, especially for the highest
stellar masses.

If the quenching of star formation, that causes the build-up of the
red galaxy sequence, is related to a {\em single} episode (i.e.\ once
the star formation is suppressed it does not restart at a significant
level), then it is highly unlikely that a direct connection between
the two phenomena -- the {\em low radio power} AGN triggering and the
star formation quenching rates -- exists. However, these results do
not exclude the possibility that low radio power AGN are responsible
for preventing the galaxies to grow higher in mass, once they have
been established as the most massive `red-and-dead' galaxies, as has
been proposed in numerous cosmological models \citep{croton06,
  bower06, sijacki06, sijacki07}. This will be further explored in
\s{sec:feedevolv} .

\section{ Implications for  galaxy formation}
\label{sec:discussion}

\subsection{ High and low radio power AGN: Different stages of galaxy evolution}
\label{sec:imact}

In the last decades studies at various wavelength regimes have
converged towards a widely accepted galaxy formation picture. Galaxies
are thought to evolve in time from an initial stage with spiral
morphology and blue optical colors towards elliptical morphologies
with red optical colors \citep{bell04, bell04dusty, borch06, faber07,
brown07, hopkins07}. This evolution is not linear but happens through
interspersed episodes of massive mass accretion onto the stellar body
as well as their central massive black holes seen as quasars
\citep{sanders96,sanders03}.  The blue-to-red galaxy evolution is
accompanied with significant build-up of stellar and BH mass; the
reddest galaxies observed in the universe are also the oldest and the
most massive ones ($M_*\gtrsim10^{11}$~\msol ; e.g.\
\citealt{baldry04, baldry06, faber07}).  Mergers between galaxies,
rich in cold gas (`wet' mergers), appear to be one of the key
processes governing this mass build-up and the blue-to-red galaxy
transformation \citep{sanders03, bell04, bell04dusty, borch06,
faber07, brown07}.  Furthermore, mergers are considered to be the
major drivers of the growth of super-massive black holes in the
centers of galaxies \citep{kauffmann00, croton06} as they enable large
amounts of gas to be funneled to, and accrete onto the central BH.

In the context of the general picture of galaxy evolution, the
following scenario for the evolution of radio AGN seems very
plausible. High radio power AGN are preferably found in galaxies that
have not yet reached the highest stellar masses and show a strong
decline with time of their number density. On the other hand, the
hosts of low power radio AGN are generally old massive `red-and-dead'
galaxies and have an almost constant space density. This implies that
the radio triggering mechanism in a galaxy is a strong function
of the host galaxy's properties and it is likely that it is linked to
different stages during the formation of massive old galaxies. In this
scenario the occurrence of the two radio modes would be naturally
linked to the overall galaxy formation process and their impact and
timescales could be observationally constrained.
 
In particular, the HERAGN phase seems to be closely tied to
major merger remnants \citep[as many hosts are found to be disturbed,
have blue colors and show signs of on-going star formation; e.g.][]
{heckman86,baum92}, as well as significant black hole growth
(supported by high -- $\sim$~Eddington -- accretion rates observed;
see \s{sec:props} ).  These major mergers are critical for the
formation of massive ellipticals. On the other hand, the LERAGN phase
is mainly observed in the most massive galaxies, where a mechanism is
required \citep{croton06, bower06} which prevents intra-cluster gas
from cooling onto the massive host galaxy. The properties of LERAGN
such as lower amounts of gas and dust compared to HERAGN
\citep{dekoff96,leon01,muller04,evans06,emonts08}, a modest BH growth
\citep{evans06, hopkins06, cao07, hardcastle07}, and fully built up
stellar and BH masses already at $z\sim1$ (\s{sec:props} ,
\citealt{kauffmann08}), support this interpretation.

In the context of the generally adopted red galaxy formation scenario,
our results thus suggest that high-excitation radio AGN (i.e.\ high
radio-power AGN; \lum~$\gtrsim10^{25}$~\wh ) occur in a stage of
galaxy evolution presumably after a gas-rich merger which induces a
phase of significant black hole growth. This active radio phase may
last $10^6 - 10^7$~yr (e.g.\ Alexander \& Leahy 1987, Shabala et
al.\ 2008). Within the post-merging process the accumulated cold gas
will be consumed by the host galaxy via star-formation and/or
black-hole accretion. Low-excitation radio AGN (i.e.\ low radio power
AGN; \lum~$\lesssim10^{25}$~\wh ) then represent a later stage of the
galaxy formation process, when the amount of cold gas in the host has
already been reduced. In this phase the galaxy's BH is accreting
  quiescently in a radiatively inefficient way, i.e.\ below a certain
  critical accretion rate in Eddington units, and most of the released
  energy is in kinetic form \citep[see e.g.][]{hardcastle07,
    merloni08}. This radio phase lasts a few times $10^7$~yr
to a few times $10^8$~yr \citep[e.g.][]{mcnamara05,nulsen05}.  Thus,
our results imply that high radio power AGN appear in an early stage,
while low radio power occur in a late stage in the evolution of
massive galaxies.

\begin{figure}[h!]
\includegraphics[bb =  54 360 486 792, width=\columnwidth
%]{intLD_AGN_z3.ps} 
]{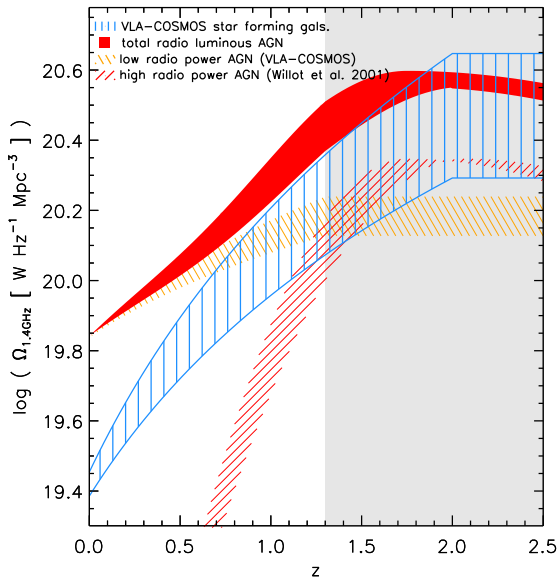}
\caption{ Evolution of the comoving 20~cm integrated luminosity
  density ($\Omega_\mathrm{1.4GHz}$) for {\em all} radio AGN (filled
  red curve) and star forming galaxies (vertically blue hatched curve)
  since $z=2.5$. The evolution of $\Omega_\mathrm{1.4GHz}$ for the
  total AGN population is a superposition of the evolution of low
  (VLA-COSMOS; orange hatched curve) and high (model taken from
  \citealt{willott01}; red hatched region) radio power AGN. The
  light-gray shaded area of the plot shows the redshift range where
  the evolution of the VLA-COSMOS AGN and star forming galaxies has
  been extrapolated ($z>1.3$; see text for details).
  \label{fig:ldintegz3}}
\end{figure}

The evolution of the integrated luminosity density, shown in
\f{fig:ldinteg} , can be well explained in our scenario.  This diagram
can be interpreted as the evolution of $\Omega_\mathrm{1.4GHz}$ for two radio
luminous, but fundamentally different, types of galaxies where the
HERAGN and LERAGN reflect modes of vigorous and modest BH
growth, respectively.

The luminosity density of the HERAGN equivalent to a mode of vigorous
BH growth evolves dramatically since $z=1.3$, consistent with e.g.\
the evolution of optically and X-ray selected quasars (see Willott et
al.\ 2001; see also Boyle et al.\ 2000 and Ueda et al.\ 2003). The
state of modest black hole growth (in our case the LERAGN) evolves
significantly slower, and it dominates the total comoving AGN radio
power output at low redshifts ($z\lesssim0.7$). At these redshifts the
most massive, `red-and-dead', often central cluster galaxies
associated with radio sources, are already in place (e.g.\
\citealt{blakeslee03}) thus providing an optimal environment for a
self-regulating process of modest BH growth (at sub-Eddington
accretion rates) and low radio power AGN activity.

 In summary, {\em the evolution of a radio selected AGN population
   seems to be consistent with the possibly bimodal evolution of black
   hole accretion modes (modest vs.\ vigorous) that occur during
   different stages in the process of the formation of (massive)
   galaxies.}

\subsection{ The co-evolution of star formation and radio AGN activity in
  radio luminous galaxies }

Black hole growth, especially merger driven growth, is expected to be
linked to starburst activity \citep{sanders96, boyle98,
  franceschini99, sanders03, croton06}.  Thus, in \f{fig:ldintegz3}
\ we compare the evolution of the integrated comoving 1.4~GHz
luminosity density for the VLA-COSMOS star forming galaxies (see S08b
for details) and the total radio AGN population, likely reflecting two
separate modes of black hole accretion, as argued above. For the
latter, $\Omega_\mathrm{1.4GHz}$ has been obtained by co-adding the
contributions of the low-power (VLA-COSMOS) and the high radio-power
AGN (\citealt{willott01}; see \f{fig:ldinteg} ).

As the peak of the cosmic star formation and AGN activity occurs at
$z>1$ \citep{hopkins04}, we extrapolate $\Omega_\mathrm{1.4GHz}$ to
$z=2.5$ as follows. For the high-power AGN we use the
\citet{willott01} model which is well constrained by 3CRR, 6CE,
  and 7CRS data at $z<3$ (see Fig.~1 in \citealt{willott01}). As can
  be seen from \f{fig:lffull} , the full (weak plus powerful radio
  AGN) Willott et al.\ model describes well also the evolution of weak
  radio AGN, at least at $z<1.3$. Here we assume that this is the case
  also for $z>1.3$. Thus, in order to be consistent with this model,
  in which the low-power AGN population is assumed to stop evolving
  beyond a certain redshift, we assume that the comoving volume
  densities of VLA-COSMOS AGN remains constant for z$>$1.3. Note
    however that this may not necessarily be realistic if the comoving
    volume density of low-luminosity AGN exhibits a turn-over, but at
    lower redshift compared to powerful radio sources
    \citep[e.g.][]{waddington01}.
%Nonetheless, combining the evolutionary trend for low-power AGN with
%the high-power AGN model reproduces the observed evolution of {\em
%all} radio AGN well, as has been demonstrated in \citet{willott01}.

To extrapolate $\Omega_\mathrm{1.4GHz}$ for the star forming galaxies
we follow Hopkins (2004) who showed, based on a compilation of
  numerous studies based on independent data sets, that ceasing the
evolution beyond $z=2$ (given the evolving radio luminosity function
for star forming galaxies) reproduces the observed cosmic star
formation history well (see e.g.\ Fig.~1 in Hopkins 2004). Thus,
  our extrapolation of $\Omega_\mathrm{1.4GHz}$ beyond $z=1.3$ is
  fairly robust as it is based on independent results drawn from
  various data sets that constrain these cosmic times well.

The integrated comoving 20~cm luminosity density for radio luminous
star forming and AGN galaxies, shown in \f{fig:ldintegz3} , appears to
evolve coevally at high redshifts ($0.7\lesssim z<2.5$), where
powerful AGN significantly contribute to the $\Omega_\mathrm{1.4GHz}$.
Both seem to flatten in the redshift range of about $1.5-2$.
Although not conclusive, this is suggestive of a link between the
process of star formation and radio AGN activity (see also e.g.\
\citealt{boyle98, franceschini99}.  This can be understood in the
scenario of galaxy formation where gas-rich mergers of spiral galaxies
govern the formation of gas-poor elliptical galaxies. As the galaxies
merge, the bulk of their cold gas, that was originally distributed
throughout the disk of the merging constituents, is funneled into the
inner few kpc of the merging system and fuels both star formation and
AGN activity (possibly sequentially; \citealt{sanders03}). Thus,
co-evolution of star forming and powerful radio AGN activity is
predominantly expected in systems with significant amounts of cold
gas, consistent with the properties found in high-excitation radio
AGN.

On the other hand, the evolution of the integrated comoving 20~cm
luminosity density for star forming and radio AGN galaxies decouples
below $z\sim0.7$ (see \f{fig:ldintegz3} ) where low radio power AGN
with modest BH growth dominate the total AGN radio-power output in the
universe (see \f{fig:ldinteg} ).  In the context of our
high-to-low-radio-power transition scenario proposed in the previous
section, this AGN activity is triggered in a later evolutionary stage
of a massive galaxy (corresponding to LERAGN), where a large fraction
of the gas reservoir has already been consumed. Thus, a close link
between star formation and AGN activity is not expected in this
regime. Such an expectation is also supported by our results on the
quenching of star formation in the context of the build-up of the red
galaxy sequence, derived in \s{sec:sfq} . We have shown that processes
causing the star formation quenching are not likely to be also
governing the triggering of low radio power AGN phases.  If they are
responsible for the decline in the star formation rate density since
$z\sim1$ (at least at the high stellar mass end; see e.g.\ S08b), then
no connection between $\Omega_\mathrm{1.4GHz}$ for low-power radio AGN
and star forming galaxies is expected.

We would like to stress that the radio energy injection into the
surrounding medium of such quiescent modes of black hole accretion has
been invoked in numerous cosmological models \citep[e.g.\
][]{croton06, bower06, sijacki06, sijacki07} as the main heating
mechanism responsible for preventing further star formation in the
given host galaxy (assuming that every massive red galaxy undergoes a
phase of radio AGN activity). This process was the key ingredient that
allowed a good reproduction of e.g. the high mass end of the galaxy
stellar mass function in the models. To shed light on the plausibility
of such a process we explore in the next section the injection
energies of our VLA-COSMOS AGN, and compare them to the recipes used
in cosmological simulations.

\subsection{The evolution of radio AGN feedback in massive galaxies}
\label{sec:feedevolv}

In this section we derive the mechanical energy injected into the
surrounding medium by low-power (VLA-COSMOS) radio AGN, which are
possible candidates for preventing star formation in massive galaxies
as proposed in recent cosmological models \citep{croton06, bower06,
  sijacki06, sijacki07}, and defined as 'radio mode' (see
\citealt{croton06} for details).

Monochromatic radio luminosity is not a direct indicator of the
mechanical energy output of a radio galaxy into its surrounding
medium. The latter can be estimated making use of the observed
interplay in galaxy clusters between radio galaxies and the hot X-ray
emitting intra-cluster gas, where radio jets induce cavities in the
hot gas and inflate buoyantly rising bubbles \citep[see e.g.\
][]{birzan04, allen06, birzan08}. As the total energy within a bubble
is the sum of its internal energy and the PdV work done by its
inflation, the mechanical radio luminosity is given by the ratio of
this total energy and the age of the bubble (see e.g.\
\citealt{birzan04, allen06, birzan08}). Using mechanical luminosities
derived by \citet{birzan04} for a sample of 18 radio galaxies in
clusters or groups, Best et al.\ (2006) have found that the relation of
mechanical and monochromatic radio power is best described by a
power-law, however with a large scatter (see Fig.~1 and eq.~2 in Best
et al.\ 2004). Based on this relation, they have computed the total
comoving volume averaged mechanical heating rate,
$\Omega_\mathrm{L\,mech}$, in the local universe provided by radio
luminous AGN, which was found to be a factor of 10-20 lower than the
local density of radio mode heating predicted in the Croton et al.\
(2006) semi-analytic model.

Here we follow the prescription of \citet{best06} to estimate the
evolution of $\Omega_\mathrm{L\,mech}$, using the most recent
calibration for the conversion of monochromatic radio to mechanical
power \citep{birzan08}. First, at each redshift we convert the radio
AGN monochromatic luminosity function (given by our best fit PLE and
PDE models; see \s{sec:evolv} ) into a mechanical luminosity density
function using eq.~16, given by \citet{birzan08}. Then we integrate it
above a minimum mechanical luminosity equivalent to
\lum~$=10^{21}$~\wh . The resulting evolution of
$\Omega_\mathrm{L\,mech}$ is shown in \f{fig:lmech} , where we compare
it to the density of the radio mode heating given in the Croton et
al.\ (2006) cosmological model. The latter has been obtained from
Fig.~3 in Croton et al., where we converted their comoving black hole
accretion rate density, $\dot{m}_\mathrm{BH}$ (shown on the right-hand
side of the y axis), into the volume-averaged heating rate utilizing
$\Omega_\mathrm{L\,mech} = \eta\, \dot{m}_\mathrm{BH}\, c^2$; where
$\eta$ is the standard efficiency of the conversion of mass into
energy ($\eta = 0.1$), and c is the speed of light (see also eq.~11 of
\citealt{croton06}).

\begin{figure}[h!]
\includegraphics[bb =  54 360 486 792, width=\columnwidth]{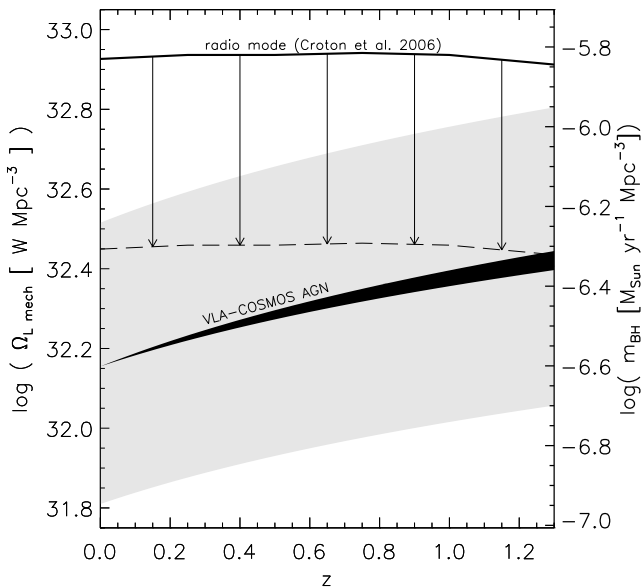}
\caption{ The cosmic evolution of the volume averaged mechanical
  heating rate $\Omega_\mathrm{L\,mech}$ for low-power VLA-COSMOS AGN,
  which are likely candidates for the 'radio mode' heating invoked in
  cosmological models (filled curve). The volume averaged accretion
  rate is shown on the right-hand side y-axis (see text for
  details). The uncertainties in the correlation between \lum\ and
  $\mathrm{L_{mech}}$ (given in \citealt{birzan08}) are illustrated by
  the light-gray shaded area.  Also shown is the evolution predicted
  in the Croton et al.\ (2006) semi-analytic model (thick line), and
  its possible value lowered by a factor of 2-3 (indicated by arrows)
  due to a systematic over-estimation of the heating rate in the model
  (see text for details).
  \label{fig:lmech}}
\end{figure}

It is important to note that both the model- and observation-based
heating rates are subject to large uncertainties. The model cooling
rates are likely to be over-estimated by a factor of 2-3 (indicated in
\f{fig:lmech} ) due to the nature of the underlying cooling flow model
(Best et al.\ 2006). On the other hand, the observationally derived
heating rates have uncertainties of the order of $\sim0.8$~dex due to
the scatter in the conversion of monochromatic to mechanical radio
luminosities.  Furthermore, they may be strongly under-estimated (up
to a factor of 6) as the work done during the inflation of a bubble
may be significantly higher than the reversible pdV work usually
assumed in the computation of mechanical luminosities (see Binney et
al.\ 2007 for details). In addition, the minimum mechanical power
  equivalent to \lum~$=10^{21}$~\wh\ has been chosen rather
  arbitrarily. For example, a lower threshold would yield a higher
  $\Omega_\mathrm{L\,mech}$. Nonetheless, the qualitative agreement
between the model and observations seen in \f{fig:lmech} \ is
encouraging for the idea that radio luminous AGN play an important
role in the process of galaxy evolution.

Based on two samples of local massive red galaxies with available
optical, radio and X-ray observations, \citet{best06} have shown that
radio source activity may indeed control the growth rate of the host
galaxy, as has been proposed in cosmological models. Best et al.\ have
found that, across a wide range of stellar masses, the radio source
heating rate agrees remarkably well with the cooling rate of the host
galaxy's surrounding hot gas halo. Averaged over time, the recurring
radio source activity can then balance the radiative losses from its
surrounding hot gas halo and thereby suppress the cooling of the gas,
and in this way control the growth rate of the galaxy \citep[see
  e.g.][]{croton06}. Here we have shown that various properties of
$z\sim1$ (low power) radio AGN are comparable to the properties of
local radio AGN. Thus it is very likely that the same radio heating --
hot gas cooling interplay, observed locally, takes place also in
$z\sim1$ radio AGN. This then strongly argues in favor of the
plausibility that (low-power) radio AGN activity (i.e.\ `radio mode')
is a key factor that prevents further mass growth in massive
`red-and-dead' galaxies across cosmic times, at least since
$\sim4.5$~Gyr after the Big Bang.

\section{Summary and conclusions}
\label{sec:summary}

Using the largest sample of low luminosity radio AGN at intermediate
redshifts available to-date drawn from the VLA-COSMOS survey, we
explored the evolution and composition of the radio AGN population out
to a redshift of z$\sim$1. In particular, we derived the radio AGN
luminosity function in several redshift bins. Compared to powerful
(\lum~$>5\times10^{25}$~\wh ) radio AGN the low radio power
(\lum~$\lesssim5\times10^{25}$~\wh ) AGN probed here show only a weak
decline in their luminosity density since $z=1.3$. 

An analysis of the host properties of weak and powerful radio AGN
shows that they form two relatively distinct classes, with powerful
radio AGN being related to modes of vigorous BH growth whereas weak
radio AGN are related to only modest BH growth. This is also reflected
in the types of host galaxies with weak radio AGN being preferentially
found in the most massive and evolved galaxies (even out to z$\sim$1).

Joint analysis of the evolution of the 20~cm luminosity density of the
AGN and the star formation rate density suggests that the powerful
radio sources are preferably triggered in major mergers following
shortly after the phase of massive star formation at high
redshifts. This would also naturally explain the strong decline of
these sources below a redshift of z=1. As the volume density of weak
radio AGN stays fairly constant, they can significantly contribute to
the heating of the their surrounding medium and thus inhibit gas
accretion onto their host galaxies as recently suggested for the radio
mode in cosmological models.

\acknowledgments The authors are grateful to P.~N.~Best for especially
deriving the SDSS/NVSS radio source fractions as a function of stellar
mass with a radio luminosity cut that matches the VLA-COSMOS survey
data.  CC acknowledges support from the Max-Planck Society and the
Alexander von Humboldt Foundation through the
Max-Planck-Forschungspreis 2005. GZ and SB acknowledge support from an
INAF contract PRIN-2006/1.06.10.08 and an ASI grant
ASI/COFIS~I/026/07/0. TP acknowledges support from PSC-CUNY grant \#
69612-00 38.

{}


\vspace{-2mm}

\begin{thebibliography}{}
\bibitem[Allen et al.(2006)]{allen06} Allen, S.~W., Dunn, R.~J.~H.,
  Fabian, A.~C., Taylor, G.~B., \& Reynolds, C.~S.\ 2006, \mnras, 372,
  21
\bibitem[Auger et al.(2008)]{auger08} Auger, M.~W., Becker, 
R.~H., \& Fassnacht, C.~D.\ 2008, \aj, 135, 1311 
\bibitem[Baldi \& Capetti(2008)]{baldi08}  Baldi, R.~D., \& Capetti, A.\ 2008, \aap, 489, 989 
\bibitem[Baldry et al.(2004)]{baldry04} Baldry, I.~K., 
Glazebrook, K., Brinkmann, J., Ivezi{\'c}, {\v Z}., Lupton, R.~H., Nichol, 
R.~C., \& Szalay, A.~S.\ 2004, \apj, 600, 681 
\bibitem[Baldry et al.(2006)]{baldry06} Baldry, I.~K., Balogh, 
M.~L., Bower, R.~G., Glazebrook, K., Nichol, R.~C., Bamford, S.~P., 
\& Budavari, T.\ 2006, \mnras, 373, 469 
\bibitem[Baldwin, Phillips \& Terlevich(1981)]{bpt81} Baldwin, J.~A.,
Phillips, M.~M., \& Terlevich, R.\ 1981, \pasp, 93, 5
\bibitem[Barger et al.(2007)]{barger07} Barger, A.~J., Cowie, 
L.~L., \& Wang, W.-H.\ 2007, \apj, 654, 764 
\bibitem[Barthel(1989)]{barthel89} Barthel, P.~D.\ 1989, \apj, 
336, 606 
\bibitem[Baum et al.(1992)]{baum92} Baum, S.~A., Heckman, 
T.~M., \& van Breugel, W.\ 1992, \apj, 389, 208 
\bibitem[Bell et al.(2004a)]{bell04} Bell, E.~F., et al.\ 2004a, 
\apj, 608, 752 
\bibitem[Bell et al.(2004b)]{bell04dusty} Bell, E.~F., et al.\ 2004b, 
\apjl, 600, L11 
\bibitem[Best et al.(2005)]{best05} Best, P.~N., et al.\ 2005, \mnras, 362, 9
\bibitem[Best et al.(2006)]{best06} Best, P.~N., Kaiser, 
C.~R., Heckman, T.~M., \& Kauffmann, G.\ 2006, \mnras, 368, L67 
\bibitem[B$\hat{i}$rzan et al.(2004)]{birzan04}
  B$\hat{i}$rzan, L., Rafferty, D.~A., McNamara, B.~R., Wise,
  M.~W., \& Nulsen, P.~E.~J.\ 2004, \apj, 607, 800
\bibitem[B$\hat{i}$rzan et al.(2008)]{birzan08}
  B$\hat{i}$rzan, L., McNamara, B.~R., Nulsen, P.~E.~J.,
  Carilli, C.~L., \& Wise, M.~W.\ 2008, ArXiv e-prints, 806,
  arXiv:0806.1929
\bibitem[Blakeslee et al.(2003)]{blakeslee03} Blakeslee, J.~P., et 
al.\ 2003, \apjl, 596, L143 
\bibitem[Bondi et al.(2008)]{bondi08} Bondi, E., et. al.\ 2008,
  accepted for publication in AJ; arXiv:0804.1706 
\bibitem[Bower et al.(2006)]{bower06} Bower, R.~G., Benson, 
A.~J., Malbon, R., Helly, J.~C., Frenk, C.~S., Baugh, C.~M., Cole, S., 
\& Lacey, C.~G.\ 2006, \mnras, 370, 645 
\bibitem[Boyle 
\& Terlevich(1998)]{boyle98} Boyle, B.~J., \& Terlevich, R.~J.\ 1998, \mnras, 293, L49 
\bibitem[Borch et 
al.(2006)]{borch06} Borch, A., et al.\ 2006, \aap, 453, 869 
\bibitem[Brown et al.(2007)]{brown07} Brown, M.~J.~I., Dey, A., 
Jannuzi, B.~T., Brand, K., Benson, A.~J., Brodwin, M., Croton, D.~J., 
\& Eisenhardt, P.~R.\ 2007, \apj, 654, 858 
\bibitem[Brusa et al.(2008)]{brusa08} Brusa, M., et al.\ 2008, submitted
\bibitem[Bruzual \& Charlot(2003)]{bc03} Bruzual, G., \& 
Charlot, S.\ 2003, \mnras, 344, 1000 
\bibitem[Cao(2007)]{cao07} Cao, X.\ 2007, \apj, 659, 950 
\bibitem[Capak et al.(2007)]{capak07} Capak, E., et. al.\ 2007,
  ApJS, COSMOS special issue, in press
\bibitem[Chabrier(2003)]{chabrier03} Chabrier, G.\ 2003, \pasp, 115,
763
\bibitem[Chiaberge et al.(2005)]{chiaberge05} Chiaberge, M., 
Capetti, A., \& Macchetto, F.~D.\ 2005, \apj, 625, 716 
\bibitem[Clewley \& Jarvis(2004)]{clewley04} Clewley, L., \& Jarvis, M.~J.\ 2004, \mnras, 352, 909 
\bibitem[Condon(1989)]{condon89} Condon, J.~J.\ 1989, \apj, 338, 
13 
\bibitem[Condon et al.(2002)]{condon02} Condon, J.~J., Cotton, 
W.~D., \& Broderick, J.~J.\ 2002, \aj, 124, 675 
\bibitem[Croton et al.(2006)]{croton06} Croton, D.~J., et al.\ 
2006, \mnras, 365, 11 
\bibitem[Cowie et al.(2004)]{cowie04} Cowie, L.~L., Barger, 
A.~J., Fomalont, E.~B., \& Capak, P.\ 2004, \apjl, 603, L69 
\bibitem[Daddi et al.(2007)]{daddi07} Daddi, E., et al.\ 2007, 
ArXiv e-prints, 705, arXiv:0705.2832 
\bibitem[de Koff et al.(1996)]{dekoff96} de Koff, S., Baum, 
S.~A., Sparks, W.~B., Biretta, J., Golombek, D., Macchetto, F., McCarthy, 
P., \& Miley, G.~K.\ 1996, \apjs, 107, 621 
\bibitem[Donoso et al.(2008)]{donoso08} Donoso, E., Best, P.~N., 
\& Kauffmann, G.\ 2008, arXiv:0809.2076 
\bibitem[Dunlop \& Peacock(1990)]{dp90} Dunlop, J.~S., \& Peacock,
  J.~A.\ 1990, \mnras, 247, 19 
\bibitem[Emonts et al.(2008)]{emonts08} Emonts, B., Morganti, R.,
  Oosterloo, T., \& van Gorkom, J.\ 2008, ArXiv e-prints, 801,
  arXiv:0801.4769
\bibitem[Evans et al.(2006)]{evans06} Evans, D.~A., Worrall, D.~M.,
  Hardcastle, M.~J., Kraft, R.~P., \& Birkinshaw, M.\ 2006, \apj, 642,
  96
\bibitem[Faber et al.(2007)]{faber07} Faber, S.~M., et al.\ 
2007, \apj, 665, 265 
\bibitem[Fabian(1994)]{fabian94} Fabian, A.~C.\ 1994, \araa, 32, 277
\bibitem[Fanaroff \& Riley(1974)]{fr74}Fanaroff, B.~L. and Riley, J.~M., 1074,
  \mnras, 167, 31
\bibitem[Finoguenov et al.(2007)]{finoguenov07} Finoguenov, A., et 
al.\ 2007, \apjs, 172, 182 
\bibitem[Fontana et al.(2006)]{fontana06} Fontana, A., et al.\ 2006,
  \aap, 459, 745
\bibitem[Franceschini et al.(1999)]{franceschini99} Franceschini, A., 
Hasinger, G., Miyaji, T., \& Malquori, D.\ 1999, \mnras, 310, L5 
\bibitem[Haas et al.(2004)]{haas04} Haas, M., et al.\ 2004, \aap, 424,
531
\bibitem[Hardcastle et al.(2006)]{hardcastle06} Hardcastle, M.~J., 
Evans, D.~A., \& Croston, J.~H.\ 2006, \mnras, 370, 1893 
\bibitem[Hardcastle et al.(2007)]{hardcastle07} Hardcastle, M.~J., 
Evans, D.~A., \& Croston, J.~H.\ 2007, \mnras, 376, 1849 
\bibitem[Heckman et al.(1986)]{heckman86} Heckman, T.~M., Smith, 
E.~P., Baum, S.~A., van Breugel, W.~J.~M., Miley, G.~K., Illingworth, 
G.~D., Bothun, G.~D., \& Balick, B.\ 1986, \apj, 311, 526 
\bibitem[Hine \& Longair(1979)]{hine79} Hine, R.~G., \& Longair,
M.~S.\ 1979, \mnras, 188, 111
\bibitem[Hopkins(2004)]{hopkins04} Hopkins, A.~M.\ 2004, \apj, 615, 209 
\bibitem[Hopkins et al.(2006a)]{hopkins06} Hopkins, P.~F., 
Narayan, R., \& Hernquist, L.\ 2006, \apj, 643, 641 
\bibitem[Hopkins et al.(2006b)]{hopkins06b} Hopkins, P.~F., 
Robertson, B., Krause, E., Hernquist, L., 
\& Cox, T.~J.\ 2006, \apj, 652, 107 
\bibitem[Hopkins et al.(2007)]{hopkins07} Hopkins, P.~F., Bundy, 
K., Hernquist, L., \& Ellis, R.~S.\ 2007, \apj, 659, 976 
\bibitem[Huynh et al.(2008)]{huynh08} Huynh, M.~T., et al. (2008);
  accepted for publication in AJ; arXiv:0803.4394
\bibitem[Ilbert et al.(2008)]{ilbert08} Ilbert, O., et al.\ 2008,
  submitted
\bibitem[Kauffmann 
\& Haehnelt(2000)]{kauffmann00} Kauffmann, G., \& Haehnelt, M.\ 2000, \mnras, 311, 576 
\bibitem[Kauffmann et al.(2008)]{kauffmann08} Kauffmann, G., 
Heckman, T.~M., \& Best, P.~N.\ 2008, \mnras, 384, 953 
\bibitem[K{\"o}rding et al.(2008)]{koerding08} K{\"o}rding, E.~G., 
Jester, S., \& Fender, R.\ 2008, \mnras, 383, 277 
\bibitem[Lacy et al.(2004)]{lacy04} Lacy, M., et al.\ 2004, 
\apjs, 154, 166
\bibitem[Laing et al.(1994)]{laing94} Laing, R.~A., Jenkins, C.~R.,
Wall, J.~V., \& Unger, S.~W.\ 1994, The Physics of Active Galaxies,
54, 201
\bibitem[Ledlow 
\& Owen(1996)]{ledlow96} Ledlow, M.~J., \& Owen, F.~N.\ 1996, \aj, 112, 9 
\bibitem[Le Floc'h et al.(2005)]{lefloc'h05} Le Floc'h, E., et al.\ 2005,
  \apj, 632, 169
\bibitem[Leon et al.(2001)]{leon01} Leon, S., Lim, J., Combes, 
F., \& van-Trung, D.\ 2001, QSO Hosts and Their Environments, 185 
\bibitem[Lim et al.(2003)]{lim03} Lim, J., Leon, S., Combes, F., \&
  Dinh-v-Trung 2003, Active Galactic Nuclei: From Central Engine to
  Host Galaxy, 290, 529
\bibitem[Marconi \& Hunt(2003)]{marconi03} Marconi, A., \& Hunt,
  L.~K.\ 2003, \apjl, 589, L21
\bibitem[Mauch 
\& Sadler(2007)]{mauch07} Mauch, T., \& Sadler, E.~M.\
  2007, \mnras, 375, 931 
\bibitem[McLure \& Dunlop(2002)]{md02} McLure, R.~J., \& Dunlop, J.~S.\ 2002, \mnras, 331, 795 
\bibitem[McLure \& Dunlop(2004)]{md04} McLure, R.~J., \& Dunlop, J.~S.\ 2004, \mnras, 352, 1390 
\bibitem[McNamara et al.(2005)]{mcnamara05} McNamara, B.~R., 
Nulsen, P.~E.~J., Wise, M.~W., Rafferty, D.~A., Carilli, C., Sarazin, 
C.~L., \& Blanton, E.~L.\ 2005, \nat, 433, 45 
\bibitem[Merloni 
\& Heinz(2008)]{merloni08} Merloni, A., \& Heinz, S.\ 2008, \mnras, 388, 1011 
\bibitem[M{\"u}ller et al.(2004)]{muller04} M{\"u}ller, S.~A.~H.,
  Haas, M., Siebenmorgen, R., Klaas, U., Meisenheimer, K., Chini, R.,
  \& Albrecht, M.\ 2004, \aap, 426, L29
\bibitem[Nulsen et al.(2005)]{nulsen05} Nulsen, P.~E.~J., 
McNamara, B.~R., Wise, M.~W., \& David, L.~P.\ 2005, \apj, 628, 629 
\bibitem[Obri{\'c} et al.(2006)]{obric06} Obri{\'c}, M., et 
al.\ 2006, \mnras, 370, 1677 
\bibitem[Owen(1993)]{owen93} Owen, F.~N.\ 1993, Jets in 
Extragalactic Radio Sources, 421, 273 
\bibitem[Sadler et al.(2002)]{sadler02} Sadler, E.~M., et al.\ 2002, \mnras,
  329, 227  
\bibitem[Sadler et al.(2007)]{sadler07} Sadler, E.~M., et al.\ 
2007, \mnras, 381, 211 
\bibitem[Sanders \& Mirabel(1996)]{sanders96} Sanders, D.~B., \&
  Mirabel, I.~F.\ 1996, \araa, 34, 749
\bibitem[Sanders(2003)]{sanders03} Sanders, D.~B.\ 2003, Journal 
of Korean Astronomical Society, 36, 149 
\bibitem[Schinnerer et al.(2007)]{schinnerer07} Schinnerer, E., et. al.\ 2007,
  ApJS, COSMOS special issue, in press
\bibitem[Schmidt(1968)]{schmidt68} Schmidt, M.\ 1968, \apj, 151, 393 
\bibitem[Schmidt et al.(1995)]{schmidt95} Schmidt, M., Schneider, 
D.~P., \& Gunn, J.~E.\ 1995, \aj, 110, 68 
\bibitem[Scoville et al.(2007)]{scoville07} Scoville, E., et. al.\ 2007,
  ApJS, COSMOS special issue, in press
\bibitem[Sijacki \& Springel(2006)]{sijacki06} Sijacki, D., \&
Springel, V.\ 2006, \mnras, 366, 397
\bibitem[Sijacki et al.(2007)]{sijacki07} Sijacki, D., Springel, 
V., di Matteo, T., \& Hernquist, L.\ 2007, \mnras, 380, 877 
\bibitem[Silverman et al.(2008)]{silverman08} Silverman, J.~D., et 
al.\ 2008, \apj, 679, 118 
\bibitem[Smol{\v c}i{\'c} et al.(2006)]{smo06} Smol\v{c}i\'{c}, V., et al.\ 
  2006, \mnras, 371, 121
\bibitem[Smol{\v c}i{\'c} et al.(2007)]{smo07} Smol{\v 
c}i{\'c}, V., et al.\ 2007, \apjs, 172, 295 
\bibitem[Smol{\v c}i{\'c} et al.(2008a)]{smo08a} Smol\v{c}i\'{c}, V., et al.\ 
  2008, accepted for publication in ApJS, arXiv:0803.0997; S08a
\bibitem[Smol{\v c}i{\'c} et al.(2008b)]{smo08b} Smol\v{c}i\'{c}, V., et al.\ 
  2008, accepted for publication in ApJ; S08b
\bibitem[Solomon \& Vanden Bout(2005)]{sol05} Solomon, P.~M., \&
  Vanden Bout, P.~A.\ 2005, \araa, 43, 677
\bibitem[Stern et al.(2005)]{stern05} Stern, D., et al.\ 2005, 
\apj, 631, 163 
\bibitem[Tasse et al.(2008)]{tasse08} Tasse, C., Best, P.~N.,
  R{\"o}ttgering, H., \& Le Borgne, D.\ 2008, \aap, 490, 893
\bibitem[Trump et al.(2007)]{trump07} Trump, J.~R., et al.\ 
2007, \apjs, 172, 383 
\bibitem[Waddington et al.(2001)]{waddington01} Waddington, I.,
Dunlop, J.~S., Peacock, J.~A., \& Windhorst, R.~A.\ 2001, \mnras, 328,
882
\bibitem[White \& Rees(1978)]{white78} White, S.~D.~M., \& Rees,
M.~J.\ 1978, \mnras, 183, 341
\bibitem[White \& Frenk(1991)]{white91} White, S.~D.~M., \& Frenk,
C.~S.\ 1991, \apj, 379, 52
\bibitem[Willott et al.(2001)]{willott01} Willott, C.~J., 
Rawlings, S., Blundell, K.~M., Lacy, M., 
\& Eales, S.~A.\ 2001, \mnras, 322, 536 

\end{thebibliography}
\end{document}